\documentclass[12pt, preprint]{aastex}
\shortauthors{Geisler     et al.}
\shorttitle{\lq Extragalactic Abundances}

\def\feh{\ {[Fe/H]}\ }
\def\met{\ {metallicity}\ }
\def\mets{\ {metallicities}\ }
\def\cl{\ {cluster}\ }
\def\cls{\ {clusters}\ }
\def\mp{\ {metal-poor}\ }
\def\mr{\ {metal-rich}\ }
\def\pop{\ {population}\ }
\def\pops{\ {populations}\ }
\def\gtsim{\ {\raise-0.5ex\hbox{$\buildrel>\over\sim$}}\ }
\def\ltsim{\ {\raise-0.5ex\hbox{$\buildrel<\over\sim$}}\ }
\def\alp{\ {$\alpha$}\ }
\def\alps{\ {$\alpha$'s}\ }
\def\Glx{\ {Galaxy}\ }
\def\Gal{\ {Galactic}\ }
\def\gal{\ {galactic}\ }
\def\gals{\ {galaxies}\ }
\def\abu{\ {abundance}\ }
\def\abus{\ {abundances}\ }
\def\MW{\ {Milky Way}\ }
\def\form{\ {formation}\ }
\def\oh{\ {outer halo}\ }

\begin{document}

\title{Chemical     Abundances and Kinematics in Globular Clusters and
Local Group Dwarf Galaxies and \\
Their Implications for 
%Formation Theories of the Galactic Halo \altaffilmark{1}}
Formation Theories of the Galactic Halo }

\author{Doug Geisler}
\affil{Grupo de Astronomia, Departamento de Fisica, Universidad de Concepci\'on, Casilla 160-C,
Concepci\'on, Chile; dgeisler@astro-udec.cl}

\author{George Wallerstein, \altaffilmark{1}}
\affil{Astronomy Department, University of Washington, Seattle, WA 98195
USA; wall@orca.astro.washington.edu}

\author{Verne V. Smith}
\affil{National Optical Astronomy Observtory, P.O. Box 26732, Tucson, AZ  
85726 USA; vsmith@noao.edu}

\author{Dana I. Casetti-Dinescu}
\affil{Department of Astronomy, Yale University, P.O. Box 208101, New
Haven, CT 06520 USA; dana@astro.yale.edu}

%\altaffiltext{1}{Invited Review}

\altaffiltext{1}{derived in part from the Henry Norris Russell Lecture delivered at the
Seattle Meeting of the American Astronomical Society in January 2003}

\begin{abstract}
We review Galactic halo \form
theories and supporting evidence,  in particular  kinematics
and detailed chemical abundances of stars in some relevant
globular clusters
as well as Local Group dwarf galaxies.
Outer halo red HB \cls tend to have large eccentricities and inhabit  the area
of the Lee diagram populated by dwarf spheroidal stars,
favoring an extraGalactic origin.
Old globulars show the full range of
eccentricities, while younger ones seem to have preferentially high
eccentricities, again hinting at their extraGalactic origin.
However, the three outer halo 2nd parameter
clusters with well-determined orbits indicate they come from three 
independent systems.
We compare detailed \abus of a variety of elements between the halo and all
dwarf galaxies studied to date, including both dwarf spheroidals and irregulars.
The salient feature is that halo \abus are
essentially unique. In particular,
the general \alp vs.
\feh pattern of 12 of the 13 \gals studied are similar to
each other and very different from the \MW. Sgr appears to be the only possible
exception.
At the \mp end the extraGalactic sample is only slightly deficient compared to
the halo 
but begins to diverge by \feh$\sim -2$
and the difference is  particularly striking for stars with
\feh$\sim -1$.
Only Sgr, the most massive dSph, has some
stars similar in \alp    \abu to \Gal stars at intermediate \mets, even the most extreme low \alp subset most likely to have
been accreted. It appears
very unlikely that a significant fraction of the \mr halo could have come
from disrupted dSphs of low mass.
 However, 
at least some of the \mp halo  may have come from
typical dSphs, and a portion of the
intermediate \met and \mr halo
may have come from very massive systems like Sgr.
This argues against
the standard hierarchical
galaxy formation scenario and
the Searle-Zinn paradigm for the formation of the \Gal halo via accretion
of ``fragments" composed of stars like those we see in
typical present-day dSphs.
The chemical differences between the dwarfs and the halo are
due to a combination of
a low star formation efficiency 
and  a high galactic wind efficiency in the former.
AGB stars are also more important in the
chemical evolution of the dwarfs.
The  \form problem may be solved if
the majority of halo stars formed within a few, very massive
satellites accreted very early.
However, any such satellites
must either be accreted MUCH earlier than postulated, before the onset of SNe Ia,
or star formation must be prevented 
to occur in them 
until only shortly before they are accreted. 
The intrinsic
scatter in many elements, particularly the \alps, indicates that
the halo was also     mixed on a surprisingly 
short timescale, a further problem for hierarchical \form theories.

\end{abstract}

\keywords{Invited Review}
%\keywords{galaxies: dwarf; galaxies: individual 
%(Sculptor); galaxies: abundances; galaxies: Local Group; globular clusters
%} 

\clearpage

\section{Introduction}
 
To terribly distort a famous phrase: our Galaxy exists, therefore it formed. 
But how? When?  What fossils can we find that might yield clues to this 
process and what means do we have at our disposal to interpret these clues?
What theories have been formulated to explain these clues and how do these
theories hold up to current evidence? We review current Galactic halo \form
theories and supporting evidence,  in particular   clues derived from kinematics
and detailed chemical abundances of individual stars in some relevant
globular clusters
as well as Local Group dwarf galaxies.

Let us first look at formation theories. 
There are two major Grand Design scenarios for the formation of the halo of our \Glx. 
These models    can apply to the \Glx as a whole but we will concentrate
on their application to the \Gal halo.

%The mean density of a large spiral is roughly      times the mean
%density of the Universe. That is a large factor to be explained. If that density enhancement had not occurred we would not be here. 
%There had to be density perturbations rather early in the Universe or else
%there could not have been any form of gravitational clumping. Since dark
%matter outweighs barionic matter by at least a factor of 20 the clumping of
%dark matter must have been the key to the origin of galaxies. Most theories
%suggest that there were vast numbers of small clumps roughly the size of
%globular clusters or small galaxies; i.e. 10$^{5}$ or 10$^{6}$ Msun. 
%In that case
%it would be necessary for 10$^{6}$ or 10$^{7}$ clumps to coagulate into a large
%galaxy. If all the clumps were of about the same size, say with a gaussian
%distribution around some mean value this would very difficult to
%accomplish. However a distribution characterized by a power law such as
%the Salpeter mass function for stars would include a low frequency of
%massive clumps which might merge and attract their small mass neightbors.
%o as the rich got richer and the poor disappeared (as some politicians would like to see happen in the United States) massive galaxies might have formed. 

\subsection{The Monolithic Collapse Model} 

In 1962 Eggen, Lyndon-Bell and Sandage (ELS) proposed a model of the \form
of the Galaxy
in which a general monolithic gravitational collapse of matter brought together 
the baryons
now observed as a spherical halo, and continued on to form the disk.
Since the infalling material had never (or
hardly ever) been processed through stars it must have been extremely
metal-poor, at least initially. They argued that the
time-scale for the collapse was short compared to a \Gal
rotation  time ($\sim 2\times 10^8$ yr), allowing for the orbital eccentricities
to vary in a potential not yet in equilibium,
but suffiently long on an evolutionary
timescale 
so that massive stars forming in the collapsing gas could live out their lives
and enrich the gas with heavy elements. 
Subsequent generations of stars had different chemical compositions, in particular
enhanced metallicity,  and 
different kinematics, as the orbits changed from very elliptical with high
energy to nearly circular, confined to the developing plane.
Hence they predicted a
gradient in metallicity and kinematics in the halo. Evidence supporting this was given     
by the correlation of orbital eccentricity, total angular momentum and
velocity perpendicular to the disk with metallicity of halo field stars.
In addition, included in the halo are especially dense stellar concentrations:
the globular clusters (GCs). If they continued to form during the collapse, then the
ELS  scenario predicts that they  should also show a metallicity gradient, 
and that they should be coeval (to within the timescale of the collapse). 
However, such  data were beyond the observational limits available at that time.
%This prediction seemed to be at least partially
%confirmed by the relatively high
%metallicity of the globular clusters in the vicinity of the Galactic Bulge
%known at that time, e.g. a \met of \gtsim solar was then accepted for NGC 
%6352 (van den Bergh 1967, Hartwick \& Hesser 1972).

\subsection{The Merger/Hierarchical Accretion Model}

Searle \& Zinn (1978 - hereafter SZ) measured and compiled the most up-to-date
and reliable Fe \abus and horizontal branch morphologies
available at the time for some 50 GCs
and investigated these properties as a function of
galactocentric distance, $R_{GC}$. Surprisingly, they found no radial
\abu gradient in the cluster system in the outer halo, with the dividing
line at about the solar $R_{GC}\sim 8$kpc. They also found
significant differences in HB morphology between
inner and outer halo GCs.

It is of interest to revisit their work with current data.
The \met scale has subsequently changed dramatically as has our ability to
measure \met and the number of clusters with reliable \mets and color-magnitude
diagrams from which to derive HB morphology now includes almost
all of the $\sim 150$ known Galactic
GCs. In addition, distances are much more robust.
In Figure 1 we plot cluster metallicity against  Galactocentric radius
 and in Figure 2
we duplicate the SZ plot of HB morphology vs. \feh for their same 4 radial
bins, using the data base maintained by Harris (2003). Here we use the 
modern quantitative definition of HB morphology:   $(B-R)/(B+V+R)$, where B,V,and R
are the number of blue, variable and red stars on the horizontal branch.
The modern data bear out the same conclusions drawn by SZ: 
beyond $R_{GC}\sim8$kpc, there is no \met gradient in the GC system. 
Secondly, the inner halo GCs show a very tight relation between \met and
HB type while the outermost halo GCs have a broad range of HB type at a given
\feh (e.g. -1.5).

%the clusters in the Galactic bulge are now considered to be of a different
%population from the halo clusters (e.g. Minniti 1995).
%Omitting those clusters there now appears
%Beyond that there are 6 clusters at distances comparable to the
%nearer dSph systems. The
%outer clusters show several characteristics that differ from the inner clusters.

There are a number of GCs in the outermost bin in Figure 2 that 
have unusually red horizontal branches despite their low metallicities,
%(with the conspicuous exception of the massive cluster, NGC 2419, at a distance
%of about 100 kpc), 
a long-standing conundrum referred to as the ``second parameter
problem" (Sandage \& Wildey      1967).  The large spread in horizontal branch 
type is best understood,
as suggested by SZ,
as a difference in age with the red HB outer halo globulars several Gyr younger than
their blue HB cousins at large radii as well as the inner halo clusters, although the solution to the second parameter
problem is still controversial  and may involve other factors such as He
\abu, internal dynamics or mass loss (e.g. Chaboyer et al. 1996, Stetson et al. 1996,
Catelan et al. 2001). Note that the inner/outer halo dichotomy now finally
appears to be well-supported by field stars as well (Carollo et al. 2007).

The lack of an outer halo GC \met gradient and the differences in HB morphology between
inner and outer halo GCs, which they interpreted as a larger age spread in the 
latter ($>1$ Gyr) than the former, led  SZ
to suggest a model in which the outer halo formed
over a longer time via the capture or accretion of external systems.
They coined the term ``fragments' for these postulated \Gal building blocks.
Zinn (1993) developed this concept further, finding that dividing the halo
GCs into 2 groups according to their HB morphology also yielded distinct
spatial distributions, rotational velocities and velocity dispersions, thus
leading to the distinction of ``old" vs ``younger" halo. 
One also speaks now of the ``dissipative" vs. ``accreted" halo in which 
the former component was formed along the lines suggested by ELS and the 
latter a la SZ (e.g. Gratton et al. 2003). These 
ideas have been thoroughly reviewed 
by Freeman and Bland-Hawthorne (2002). 
Note that
the clusters in the Galactic bulge are now considered to be of a different, 
bulge, population from the halo clusters (e.g. Minniti 1995).

%    Once the word "capture" is invoked the question is immediately raised
%as to how much of the Galactic halo may have been
%aquired by the capture of dwarf
%galaxies and/or globular clusters. The globulars could hold themselves
%together when captured (at least initially)
%but the dwarf systems would be pulled apart into individual
%stars by gradients in the gravitational field of the protoGalaxy.
%SZ noted the requirement that a large
%central mass was necessary to attract small neighboring systems.

The SZ scenario has been given a firm   cosmological footing in recent years. 
Modern cosmological theories based on the currently favored
$\Lambda$ Cold Dark Matter ($\Lambda$CDM) paradigm
posit hierarchical structure formation on all physical scales (e.g. White
\&  Rees 1978, Navarro, Frenk \& White 1997). 
They predict that all \gals,
including the \MW, form as part of a local over-density in the primordial
matter distribution and grow
via the accretion of numerous smaller building blocks
(SZ ``fragments') which themselves formed similarly. 

As originally noted by SZ and later by Zinn (1980), 
obvious candidates for these building blocks are the present day dwarf
spheroidal (dSph) and dwarf irregular (dIrr) \gals. Such \gals are the 
most numerous in the Universe and the dSphs in particular
are found to surround both the \MW 
and M31 in large numbers (although their numbers fail by an order of magnitude
or more to match those predicted by $\Lambda$CDM theories (Klypin et al.
1999). There are undoubtedly other faint dwarfs left to be found -- e.g. 
the Sloan Digital Sky Survey has turned up 
very recently a number of new dSph companions to the \MW 
(Willman et al. 2005, Kleyna et al. 2005, Belokurov 2006b, Zucker et al. 2006a,b,c) -- but it is
unlikely that the numbers required by $\Lambda$CDM theories really exist, 
although the most recent estimate (Simon \& Geha 2007) suggests that there are
only a factor of $\sim4$ too few dwarf galaxies currently known). 
Many of them are known to contain at least a sizeable fraction of
stars that are similar to those in the halo - namely old and \mp.
Graphic proof that accretion of such \gals
does indeed occur has existed ever since the 
discovery of the Sagittarius dSph (Sgr -- Ibata et al. 1994), 
which is clearly being assimilated
by the \Glx along with its coterie of GCs. Not only 
are Sgr field stars now forming part of the MW halo, but its several
GCs are also now becoming part of the Milky Way's GC system.
It is also theorized
that a significant fraction of halo field stars were once members of a GC.
Indeed, we now know that GCs do indeed disrupt - the most beautiful example
being that of Pal 5 (Odenkirchen et al. 2003). 

\subsection{The Search for Bulding Blocks: \\
Comparing Observations with Theory}

One can think of a number of observational tests to probe the predictions
of different \gal \form  theories. An obvious test is to directly compare
the stellar \pops of the surviving dSph or dIrr systems with that of the halo. 
If the halo is indeed made
up in large part by dissolved systems initially
like the dSphs or dIrrs we see today, one would expect
to find many similarities in their stellar \pops. 

One  way to do this is 
to compare the CMDs in detail and try to  set limits on the
percentage of present day
dSph populations that may have contributed to the halo. By
comparing the turnoff colors in these systems, Unavane, Wyse \& Gilmore (1996)
first set a surprisingly low upper limit of $\sim 10\%$ on this contribution, as the
intermediate-age stars generally found in dSphs are lacking in the halo.
This was the first strong 
hint that at least the present day
dSphs may not be the generic galactic building blocks they are
often imagined to be.

Another more stringent but observationally more difficult approach is a direct
comparison of the detailed chemical compositions of
stars  from the two environments, based on high resolution spectroscopy.
Clearly, if the halo formed from dSphs or dIrrs
or objects like them, their chemical
makeups should be similar. Such a study      has been dubbed ``chemical
tagging" by Freeman \& Bland-Hawthorn (2002). The catchphrase ``near field
cosmology" also applies as we are probing     cosmological galaxy \form theories using
the nearest \gals as our testbeds.
The value of such a comparison was
recognized long ago and served as one of the key science drivers
for the installation of high resolution spectrographs on the new generation
of 6-10m telescopes, as even the brightest stars in the older \pops in
even the 
nearest \gals were really beyond the reach
of high resolution spectrographs on 4m telescopes.
In the context of this review, when referring to \abus of individual elements
other than simply ``\feh'', we restrict ourselves to data with a resolution
R$\gtsim20,000$
in which the chemical abundances of individual elements in a single star have
been derived from a detailed model atmosphere analysis. 
(NB - for historical purposes, ``\met" refers to the Fe \abu,  i.e. \feh  and
we will adhere to this).
For this purpose, high S/N data ($\gtsim 50$) is certainly preferable.
Of course, spatial and kinematic as well as chemical data are needed to
really disentangle the predictions of different theories (e.g. Venn et al. 2004,
Font et al.
2006a,b). In this review we will mainly focus on high resolution \abu results
but will also discuss some recent kinematic evidence.

In addition to providing a key observational test for \gal \form models,
detailed chemical \abus of our nearest galactic neighbors also allow us
to reconstruct their own chemical evolution history, investigating the 
relative contributions over time to different elements by
SNe Ia, SNe II, AGB stars, etc.,
as well as to help constrain their
star \form history. dSphs are also believed to be local analogs to 
damped Ly$\alpha$ \gals and/or the 
many distant faint blue \gals seen in very deep images.

After the installation of instruments like UVES on the VLT and HiRes on
Keck, the field of extraGalactic chemical \abus has really taken off. 
Detailed \abus now exist for at least a few stars in each
of the dSphs associated with the \MW (except for Leo II)
as well as for several of the nearest
dIrrs, thanks in large part to the efforts of M. Shetrone, K. Venn, E. Tolstoy,
V. Hill, J. Johnson, P. Bonifacio, T. Smecker-Hane, A. McWilliam  and 
collaborators,  and of course the Magellanic Clouds, although, surprisingly,
the amount of high quality data for older
LMC stars is very limited and indeed virtually
non-existant for the SMC. 
We note that this is still very hard work: the brightest target stars are
typically $V\sim 17$ and even with good seeing on an 8m telescope and an
efficient spectrograph they require several hours to achieve adequate S/N at
high resolution.
In addition, many new studies
have been carried out on major samples of halo stars which provide  a 
much better sample of our own \Glx to compare to.
The results have had major implications on our  understanding of how our
\Glx, as well as other \gals, might have formed. 
This field will continue to expand rapidly  in the near future, especially with
the recent implementation of  multiplexing spectrographs
like FLAMES on the VLT and MOE+IMACS on Magellan.
These new instruments will further revolutionize this field yielding orders
of magnitude more stars per galaxy. Nevertheless, we feel a summary of our current
knowledge in this field is timely and hence this Review, although we fully
expect many of the details will be outdated in short order.
We generally restrict ourselves to papers published or in preprint form
by  mid 2007.

The paper is organized as follows: We start by investigating outer
halo GCs in the 
context of their possible extraGalactic origin. We next discuss 
the first galaxy known to have been (or is being) captured by the \Glx - the Sgr
dSph. We then compare detailed \abus of a variety of important elements, including
O, Mg, Si, Ca, Ti, Na, Fe, Ba and Eu in several different galaxies. We follow
with   a long discussion of the implications for \form  scenarios
of the particularly important\alp element
\abus in nearby galaxies compared to those in the halo. We close by 
enumerating some of the problems associated with the latest galaxy \form
theories as applied to our halo.

\section{Globular Clusters of the Outer Halo}

In order to test the SZ/hierarchical merger idea that at least the outer halo
was accreted from fragments, by comparing \abus in the halo with those in
potential building block \gals, we need to not only examine \abus in other
\gals but also representatives of the outer halo. We take the same working 
definition of the ``\oh" as SZ: $R_{GC}>8kpc$. Several recent studies have
explored \abus in halo field stars and GCs, e.g. Venn et al. (2004),
Beers \& Christlieb (2005). We present
a brief review of the properties of \oh GCs, including not only \abus but also
orbital eccentricities which are now available for some of these objects.

Let us first return to Figure 1 and focus on the more distant \cls. The bulk
of the \oh globulars have [Fe/H] between -1.2 and -2.3.
%show no noticable gradient. There is also a clump of clusters with [Fe/H]
%between -0.4 and -0.7 and logRg$<$1.2. They probably belong to the thick disk. We
%omit Pal 12 and Ter 7 since they are now being captured along with the
%Sagittarius Galaxy. 
Note the significant gap in the GC radial distribution in which no \cls are 
found between $\sim 40-70$ kpc (Zinn 1985).
In the far halo, beyond the gap,
 there are 6 clusters, with Galactocentric distances 
extending to more than
100 kpc, entering 
the realm  of the nearest dSphs, which (excepting Sgr) have $R_{GC}\sim 70-
90$ kpc (Mateo 1998). Five of these \cls have 
[Fe/H] between -1.4 and -1.8 while a single cluster, NGC 2419, has [Fe/H]
$= -2.14$. 
A glance at Figure 2 reveals that, of these 6 very distant \oh \cls, all
except NGC 2419 have very red HBs.

These 5 outer halo, RHB \cls -- Pal 3, 4, 14, Eridanus and AM1 -- are then excellent candidates
for accreted \cls, especially if age is the second parameter.
In order to investigate this further, it would be of great
interest to determine their kinematics and orbits.
Unfortunately, because of their great distances their proper motions are
not yet known so we do not know their orbital eccentricities, but it is 
likely that they do not have penetrating orbits or they would have lost 
sufficient members so as to no longer be recognizable, except possibly by their debris streams.
  
However, with the galactic orbits of many other
globular clusters in hand (Dinescu et al. 1999,
and further updates) we can prepare plots like Figure 2 but now 
including eccentricity.
In Fig 3 we plot the metallicity against HB type in what has been called a
 Lee  diagram (Lee et al. 1994) for GCs with known eccentricities, 
with the symbol size proportional to the eccentricity.
Two disk clusters --- 47 Tuc and NGC 6838 --- are marked with crosses, and Pal 12
is labeled.  For comparison, we also plot Sgr and its 4 main-body clusters
(filled circles) and the Fornax clusters (star symbols; for these clusters the symbol
size is not related to the eccentricity, which is unknown). In addition, we include known mean
values for the field \pops of dSphs (small triangles), with values taken from Grebel et al.
(2003) and Harbeck et al. (2001). Again, eccentricity values are not known for these
\gals. According to Lee et al. (1994), lines of equal age may be drawn
in this diagram. We have done this, using the isochrones from Lee et al. (2002).
The full range in age is about $\pm2$ Gyrs, some of which may be due to the
uncertainties in the whole process. Blue HB clusters at a given low metallicity
(or `first parameter' clusters) appear to have the full range of eccentricities, while
red HB clusters at a given low metallicity (or 2nd parameter clusters) seem to have
preferentially large eccentricities. These latter \cls also fall in the
same general region of the diagram as the general \pops of the dSphs. Both of
these facts are evidence in favor of an extraGalactic origin for the 
2nd parameter clusters.

The GCs               of Fornax range from HB = -0.4 to 
+0.5 though their apparent [Fe/H] range is only from -2.1 to -1.8. Fornax appears
to have its own second parameter problem (Smith et al. 1996)! 
For the  dSph companions of
M31 the HB values are all near -0.7 with [Fe/H] ranging from -1.5 to -2.0 which
is within the uncertainties of photometric metallicities. Why they should all be
so similar while the dSph systems surrounding our \Glx and their associated globulars are so different is an intriguing question.

The large range in age and metallicity of the globulars and dSph systems shows
the large range of conditions that must have been present as these systems 
formed. We should remember, however, that the globulars and dSph systems that we see
today are the ones that, if they were captured, have not yet been totally
assimilated into the halo. Their
stellar content as well as their orbits may differ from those that were captured
earlier in the history of our galaxy, and were primarily
responsible for contributing their stars to the current halo.   
Perhaps it should be no surprise that halo stars have compositions more like
dSph systems when their \feh values were close to -2 rather than their present
values nearer to \feh of -1 (see Section 5).

De Angeli et al. (2005) derive precise relative GC ages from
homogeneous photometric data sets: ground based and HST. We plot
the ages determined from this study
on the Carretta \& Gratton (1997) metallicity scale
as a function of orbital eccentricities in Figure 4.
The two disk clusters 47 Tuc and NGC 6838 are indicated with
star symbols in this plot. Old clusters show the full range of
eccentricities, while younger ones seem to have preferentially high
eccentricities. Two well-known 2nd parameter clusters are not present in the
age-eccentricity plot: NGC 7006 and Pal 13. These two clusters also have high eccentricities,
0.69 and 0.76 respectively. Finally, Omega Cen --- a system widely accepted to be
the nucleus of a captured dwarf elliptical galaxy ---
has a high orbital eccentricity (0.57).  More importantly, to match 
Omega Cen's  present
orbital characteristics, Tsuchyia et al. (2003, 2004) show that its progenitor
started with an orbital eccentricity of 0.9.

If indeed the majority of 2nd parameter clusters
(or the younger clusters) were
born in satellite systems of the Milky Way,
their orbit shapes indicate that these systems were rapidly destroyed due to
their penetration into the inner, denser regions of the Galaxy. Only satellite systems
with low orbital eccentricities could have survived. Fornax is such an example. It
has an orbital eccentricity of $0.27\pm 0.16$ (Dinescu et al. 2004, see however Piatek et al.
2002 for a different proper-motion determination which corresponds to an eccentricity= 0.52).
The LMC too has a low-eccentricity orbit ($\sim 0.35$), while Sgr's is rather
moderate (0.53) (e.g., Dinescu et al. 2001, 2005).

If the assumption that 2nd parameter clusters were captured is correct,
one can envision testing the hypothesis that the outer halo was assembled
from just a few massive satellites, as recent models predict (Robertson et al.
2005 -- see Section 5).
Specifically, by quantifying the amount of phase-space association of distant globular clusters ---
once accurate proper motions are determined for a good number of these clusters, e.g. from
the SIM/PlanetQuest or GAIA projects --- one can
set constraints on the generic number of SZ fragments/satellites that made up
the halo. We note here an initial attempt along these lines:
There are three 2nd parameter
clusters with well-determined orbits: Pal 13, NGC 7006 and NGC 5466. 
These 3 clusters have high-energy orbits (i.e. they belong to the outer halo),
and their orbits are highly eccentric. They are thus good candidates 
to have been produced in the SZ fragments which were later captured and
disrupted. Their orbits however indicate that they come from {\bf three independent
systems}.  If the \oh was indeed made out of only a few very massive systems,
one would expect better phase-space association than demonstrated by these
\cls but of course the numbers are still very small.

\section{The Sagittarius System}

     Ibata et al. (1994) noted a large number of stars of magnitude about 17 and
fainter in a direction towards the Galactic center
that appeared to be at about the same distance and radial velocity. 
The Sgr system is now known to extend over many degrees and to show multiple
 tidal
tails that are    extremely long (Belokurov et al. 2006a). In additon to the main
body of the Sgr system there are 4 nearby globular clusters that appear to be
co-moving with Sgr and at a similar distance. These are M54, Terzan 7, Terzan 8
and Arp 2 (Da Costa \& Armandroff 1995). 
The relatively young globular cluster Pal 12, somewhat further away, seems also to be co-moving with the 
system (Dinescu et al. 2000) and is now generally believed to be a Sgr member
(Cohen 2004, Sbordone et al. 2006), as is Whiting 1 (Carraro et al. 2007).
M54 is one of the most massive globulars in the Galaxy and was probably
the nucleus of Sgr when it was a more independent entity than it is now, 
possibly a nucleated dwarf galaxy (Sarajedini \& Layden 1995).
Two properties of these globulars are very revealing of the star formation
and evolution of small systems -              age and chemical composition.

\subsection{Ages and Overall Metallicities in Sgr and its Globular Clusters}

%     In figures 2-7 we show the color-magnitude diagrams (CMDs) for Sgr, M54,
%Ter7, Ter8, Arp2,and Pal12. The references from which each figure has been pirated are given in the figure captions. 
To first order CMDs        provide
information on both the age and metallicity of a globular cluster. Ages can
be estimated by the luminosity of the turn-off. The estimate of a system's 
age can be complicated by the presence of stars of either several discrete
ages or a range of ages and/or compositions.
Once an age for a component of a system has been
established, a rough idea of its overall metallicity can be estimated from the absolute 
magnitude (and color) of the
brightest red giants. These procedures have been known for half a century
(Sandage, 1953; Hoyle and Schwarzschild, 1955), and greatly refined by many
authors over the years. Our best current estimates of the ages and metallicities
(based on both photometric and spectroscopic studies) of Sgr and its globular
clusters are:
Sgr: 1-13 Gyr, -1.6 -- +0.1; M54: 13 Gyr, -1.55; Terzan 8: 13 Gyr, -2.0;
Arp 2: 11 Gyr, -1.8; Terzan 7: 7.5 Gyr, -0.6; Pal 12: 6.5 Gyr, -0.8;
Whiting 1: 6.5 Gyr, -0.65 (taken mostly from Harris 2003).
  The range of ages and metallicities both
within Sgr and its clusters is remarkably large. 
Sgr itself seems to have stars with a range of ages from 12 or 13 Gyrs
down to about 1 Gyr with metallicities that range from [Fe/H]$\sim -1.5$ to
solar    (Layden \& Sarajedini 2000, Smecker-Hane \& McWilliam 2002, Bonifacio
et al. 2004, Monaco et al. 2005). 
The globulars do not show internal spreads of age or metallicity, but
show large differences in these quantities when compared with each other. 
M54, the likely nucleus of Sgr, is 13 Gyr
old with  \feh = --1.55 (Brown, Wallerstein, \&  Gonzalez 1999).
Terzan 8 is as old
as Sgr and M54 and even more metal-poor. Arp 2 is similarly \mp but likely a
few Gyr younger. On the other hand, Terzan 7,    Pal 12 and Whiting 1 are many Gyr 
younger and much more \mr.
These  are three of the few definitively young globulars of the halo. 
The positions of Terzan 7 and Pal 12 in the Lee diagram are shown in Figure 3. 
Despite their wide range in ages and \mets, 
the Sgr systems appear
to fall along the $\Delta t = 0$ Gyr line. However Ter 7 and Pal 12
could have any age from 0 to -4
Gyr while Arp 2 and Ter 8 could lie anywhere from -2 to +2 Gyr. As Fig 3
shows, clusters with (B-R)/(B+V+R) near -1 and +1 cannot have their ages
determined from their position in Fig. 3.
%all of these systems lie very close to
%the $\Delta t=0$ Gyr line, presumably of similar age to the old halo GCs.

%The real mystery is how the Sgr system could give birth to 2 metal-rich
%clusters several Gyrs after its own primary star formation period. Arp 2 is
%metal-poor with a red horizontal branch. If its horizontal branch indicates  
%youth the problem becomes worse because we need to explain how  Sgr could
%have assisted in the production of a metal-poor cluster at about the same
%time as Pal 12 and Ter 7 were forming. 

These wide ranges of age and metallicity raise significant questions 
regarding their past histories. It is unclear how Sgr, a dSph or perhaps
originally a nucleated dE
galaxy, could have retained sufficient interstellar matter for continuous or
intermittent star formation over a period of 10 Gyrs and cluster \form over
6.5 Gyr, half of its lifetime.  Supernova ejecta 
cannot be retained by the gravitational field of a small galaxy (although see
Marcolini et al. 2006), but perhaps Sgr was above the minimum mass required.
The only non-gravitational
way to retain such high velocity material is by running it into ambient
interstellar matter whereby its kinetic energy can by converted to thermal
energy in a shockwave and then radiated away. Even that process will not
work for ever as momentum transfer to the interstellar clouds will eventually
blow them out of the system. We just do not know how such systems could
have ``reinvented themselves" every few Gyrs. This problem pervades the dSph 
systems (Mateo 1998, Table 6).
%In addition it is very difficult
%to imagine how Sgr could either have accumulated or generated globular
%clusters with a wide range of ages and metallicities.
However, this problem is actually least severe for Sgr, the most massive dSph.
      
\subsection{Comparing Chemical Abundances in Sagittarius and its Globulars}

To treat   Sgr and its globulars as a single evolving system, we plot the mean
[X/Fe] values for elements X against [Fe/H] in Figure 5 for M54, Pal 12, Ter 7, and Sgr
itself (5, 4, 3 and $\sim 25$ stars, respectively). 
There is no high resolution data for Ter 8,   Arp 2 or Whiting 1 as of this writing.
The data are from
Brown et al (1999), Cohen (2004), Tautvaiseine et al. (2004), Smecker-Hane and 
McWilliam (2002) and Bonifacio et al. (2004). 
For Na/O     with a likely significant range within one of the systems we use the high end for the O and the low end for the Na to handle possible internal depletion of O and enhancement of Na.

   Most species do not show a \met trend. The mean values of [X/Fe] are approximately
--0.25 for Na, +0.1 for O, 0.0 for the 
$\alpha$-elements (Mg, Si, Ca and Ti), --0.15 for the light s-process elements
Y+Zr, and +0.5 for the r-process element Eu. For the heavy s-process elements 
Ba+La there is a well established trend from near 0.0 at [Fe/H]=-1.5 to
+0.4 at [Fe/H]=--0.5. A similar though stronger trend in the heavy s-process elements has been seen in Omega Cen 
(Vanture, Wallerstein, \& Brown 1994  ; Norris \&  Da Costa 1995 ).
%For oxygen there is a downward trend from [O/Fe]=0.25 near [Fe/H] = -1.5 to
%-0.2 at [Fe/H]=0.0. That trend is clearly seen in Sgr (McWilliam and Smecker-Hane   ). We discuss this further in the next section. 

In Figure 6 we show the mean abundances of various species in each of the Sgr
system's globulars and for the most metal-rich stars in Sgr itself. For the 
abscissa we use the age assigned to each system.
Starting with oxygen we see that [O/Fe] has remained constant from the time of 
star formation in M54 until recent star formation in Sgr. The slightly discrepant
value for Ter 7 may well be due to random errors when the O abundance depends
on only one line (which must be corrected for atmospheric absorption) in a few
stars. Note that Sbordone et al. (2006) have derived a slightly lower mean 
value of 0.18 from 4 stars.
[\alp/Fe], which is determined more reliably, descends from +0.2 to
0.0 or -0.1 from 13.5 Gyrs to about 2 Gyrs ago. Evidently the ratio of SNe II's
to SNe Ia's did not change very much during that interval.
For Sgr and  its globular clusters
[\alp/Fe] is low compared to the halo and nearly independent of [Fe/H] (but does
show an indication of a small age gradient).
We will discuss this
further in Section 5.
Turning to [Na/Fe] we see that it descends from +0.1 to -0.4 as Fe built up more
rapidly than did Na. The light s-process elements, represented by Y and Zr
do not show a significant trend; but the difference between Ter 7 and Pal 12
is intriguing, as it is for Na. The heavy s-process elements, Ba and La, show
a strong increase with time, while the r-process species, Eu, has 
remained high -- near
[Eu/Fe] = +0.5 - for all systems. Both the heavy s and r-process patterns are similar
to what is seen in many halo stars.
%In Table 2???,  we compare the [\alp/Fe] value in the Sgr system and its globulars
%with halo stars, globular clusters, and damped Lyman-alpha (DLA) systems. The
%parameter is metallicity rather than age. Of course the halo stars and the 
%globular
%clusters are in agreement. 
%Interestingly the
%DLAs from [Fe/H] = -3.0 to -1.5 appear to be similar to the halo stars and the 
%globular clusters. However Fe and
%all of the \alp elements except S
%are subject to depletion onto grains in the interstellar medium. Hence they will
%be reduced    in the gas phase of the DLAs but will be incorporated into
%stars when the DLA gas condenses into stars. However S and
%the iron-peak element, Zn, tend not to be captured onto grains. They have been
%observed in DLAs from [Zn/H] = -2.0 to -0.5 and show low values, rather like the
%dSph systems and the Sgr system.
%It is surprising that the DLAs show little dependence of
%metallicity on redshift.

\section{Detailed Chemical Evidence from Extra-Galactic Systems}

\subsection{Chemical Tagging     }

Detailed abundance distributions containing strategic elements
are very useful as a tool for comparing the chemical compositions of the
dwarf galaxies with the compositions of Galactic halo, thick disk, and
thin disk stars.  The elemental abundances of Galactic populations have
been rather well-studied for many key elements; these abundances are
available and well-sampled from solar metallicities down to [Fe/H]
of about -4.0.  Galactic studies that are used as comparisons
here are Edvardsson et al. (1993) and Reddy et al. (2003) for the thin disk,
while Prochaska et al. (2000) and Reddy et al. (2006) provide abundance distributions for
thick disk stars.  At lower metallicities, recent fairly large surveys
of abundances in halo stars include Fulbright (2000), Johnson (2002),
Fulbright \& Johnson (2003) and Cayrel et al. (2004).  
The work by McWilliam et al. (1995)
is useful for including stars having extremely low metallicities
([Fe/H]$<-3.0$).  This section will focus on a comparison of field
stars, however there are a number of globular clusters in the metallicity
range from [Fe/H] = -1 to -2.
Below [Fe/H]= -2 the number of halo stars diminishes steadily toward
[Fe/H= -4 with several intriguing objects that shows $[Fe/H] < -5.0$.
Only a few globulars have $[Fe/H]<-2$ and none are known below -2.5.

Abundances in some dwarf galaxies of the Local Group
will be compared to Galactic abundances from the
studies described above.  In particular, the studies by
Shetrone et al. (1998, 2001, 2003) and Geisler et al. (2005) of dwarf spheroidals
will be used, as well as the LMC studies by Smith et al. (2002),
Hill et al. (2000), and Korn et al. (2002), and the Sgr galaxy by
Smecker-Hane and McWilliam (1999), 
McWilliam et al. (2003) and McWilliam and Smecker-Hane (2005a).
Mateo's (1998) Table 6 shows [Fe/H] values for old
populations of 28 Local Group \gals   ranging from -2.2 to -1.0. The distribution
is similar to that of the Galactic globulars with a median at -1.55.
All  of the Local Group
galaxies have a range of metallicities and virtually all also have
a range of ages, a
phenomenon seen for certain amongst the GCs only in $\omega$ Cen,
which is a likely capture. Several recent papers have shown
that the dwarf systems show significant ranges not only in [Fe/H] but also
in the ratios of various elements to Iron (Shetrone et al. 2003; Geisler et
al. 2005; ).
The comparison of Galactic and dwarf galaxy populations will rely
primarily on how the ratios of various elements to iron
(expressed as [X/Fe]) depend on the overall metallicity, as defined
by [Fe/H]. These ratios reveal details about how different galactic
systems enriched themselves as the overall metallicity
increased.

Our comparison begins by investigating the behavior of oxygen, a
quintessential product of massive star enrichment and dispersal
via SNe II that tracks the formation of massive stars.
Figure 7 shows the behavior of [O/Fe], as determined from either
the 6300\AA\ line, or the infrared vibration-rotation OH lines, for
the Milky Way, the Large Magellanic Cloud (LMC), the Sagittarius
dwarf galaxy, and Sculptor (hereafter Scl).  This first comparison of [O/Fe] is restricted
to these three satellite galaxies as these are the ones that have had
the largest numbers of member stars analyzed.

The top panel in Figure 7 summarizes results for stellar members
of the Milky Way thin disk, thick disk, and halo.  The general
structure of [O/Fe] versus [Fe/H] for Milky Way populations
is now reasonably well-defined; values of [O/Fe] are larger for the
metal-poor halo stars with typical values being +0.5 for
[Fe/H]$\le$-1.0.  There may be a slight slope such that [O/Fe]
increases with decreasing [Fe/H], but such a slope if present is
not large (being less than about 0.10 dex per dex).  At
[Fe/H]$\sim$-1.0, [O/Fe] begins to decrease and reaches a value of
0.0 at about solar [Fe/H].  The elevated values of [O/Fe] for \mp stars can be
understood as both O and Fe being produced as a result of SNe II
with the masses being distributed along a standard IMF (e.g.,
yields from Woosley \& Weaver (1995) would predict
[O/Fe]$\sim$+0.3 to +0.6, with uncertainty due primarily to
the mass cut for Fe production), while the decreasing [O/Fe] for higher 
\mets occurs as SNe Ia inject mainly Fe into the ISM.

%The downturn in [O/Fe] starting at $\feh
%\sim -1$ is
%believed to result from nucleosynthesis from SNe Ia, which produce
%large amounts of Fe, but take much longer to evolve than the
%massive stars that end as SNe II.  The value of [Fe/H] at which
%[O/Fe] begins to decline depends primarily on how efficiently
%gas is turned into stars in a particular galactic system; if
%star formation is rapid and efficient (that is a large fraction
%of gas is turned into stars) then the metallicity as defined
%by [Fe/H] increases on timescales short compared to SNe Ia
%evolution ($\gtsim10^{9}$ yr).  By the time that SNe Ia
%begin to produce iron and drive [O/Fe] to lower values, the
%overall metallicity of the system is relatively large.  In

%In addition, loss of processed stellar ejecta (via a galactic wind),
%or infall of matter, can influence the shape of the
%[O/Fe]--[Fe/H] relation, although the primary effect results
%from the star formation efficiency.  
The solid curve in the top
panel of Figure 7 is a simple model of chemical evolution for
the Milky Way in which oxygen and iron from SNe II and SNe Ia are 
added to gas.  
This model                          is taken from the
paper by Smith et al. (2002) in their study of the
LMC.  The model is described in their paper, but is
simply a numerical model in which yields of O and Fe
from SNe II and SNe Ia are added at certain rates into a
mass of gas that is undergoing continuous star formation.
The numerical model from Smith et al. is similar in its
assumptions to the analytical models of Pagel and
Tautvaisiene (1995).
In this model oxygen yields are taken from
Woosley \& Weaver (1995) and convolved with a Salpeter mass
function.  Based on results from Timmes, Woosley, \& Weaver (1995),
the Fe yield from SNe II was set to 0.15M$_{\odot}$ per event;
these numbers produce a value of [O/Fe]=+0.5 from the mass-convolved
SNe II.  The downturn in [O/Fe] at [Fe/H]$\sim$-1.0
(as is observed) was produced with an average SNe II rate of
one per 100 years and SNe Ia beginning after 1.2 Gyr at a rate of
1/3 that of SNe II, with each SNe Ia event producing 0.7M$_{\odot}$
of Fe.  Of course such a model is not unique and is an
oversimplification; however, it does a fair job of fitting the
Galactic trend and this model relation can serve as a fiducial curve
in comparison to the dwarf galaxies.

The second, third, and fourth panels of Figure 7 summarize current
results for the LMC, the Sgr galaxy, and Scl, respectively.
Superimposed on each dwarf galaxy [O/Fe]--[Fe/H] relation is the
fiducial model constructed for the Milky Way.  There is a consistent
trend in each of the dwarf galaxies to exhibit lower values of
[O/Fe] relative to the Milky Way as [Fe/H] increases.  Both the
LMC and Sgr have quite similar relations, with only moderately
low [O/Fe] values compared to the Galactic curve.  Scl, on
the other hand, is strikingly different from the Milky Way, with
[O/Fe] values beginning a precipitous decline at a low overall
metallicity ([Fe/H]$\sim$-1.7).  
%As discussed above,
%decreasing values of [O/Fe] at lower [Fe/H] can be modelled by
%less efficient star formation rates.  Lanfranchi \& Matteucci (2004)
%have modelled dwarf galaxies in detail and can reproduce the
%behavior observed in Figure X by a combination of much lower
%star formation efficiencies coupled with mass loss via galactic
%winds.  Thus far [O/Fe]--[Fe/H] relations in dwarf galaxies points
%to much less efficient star formation rates over time, with perhaps
%galactic winds also playing a role in shaping the chemical evolutionary
%patterns.

We next comment on the other commonly studied \alp elements -- Mg, Si, Ca
and Ti -- in turn. Here we will concentrate on the 2 \gals with the best
existing data: Scl (Shetrone et al. 2003 - S03, Geisler et al. 2005 - G05)
and the LMC (Johnson et al. 2006 - J06, Pompeia et al. 2006 - P06). 
%In both cases, however, the sample sizes
%are still very small  -- only 9 and 10 stars respectively. 
The Scl stars were selected to cover as much of the well-known \met spread
in that galaxy as possible, and recent results (Tolstoy et al. 2001, 2004)
indicate success in this regard. As noted before, good high resolution
\abus for a variety of elements for older
MC stars are sorely lacking. The best studies to date with such
\abus are those of J05, who observed 10 giants in a total of 4 old LMC
\cls covering a range of \mets from $\sim -1.2$ to $-2.2$, and the FLAMES
study of P06, who investigate \abus of $\sim 60$ giants ranging in \met from
-1.7 to -0.3. We include here only the LMC stars that overlap in \met with 
the Scl sample.

Figure 8a-d shows the behavior for these 4 elements.
Note that typical halo \abus are $\sim +0.4$ for each element over this \met
range.
In general,                        the Scl and LMC stars are significantly
depleted in
all four \alps with respect to their \Gal counterparts at similar \met, as 
already noted by S03, G05, J06 and P06.
The exceptions tend to be the more \mp stars which tend to be very similar
to the halo or only slightly depleted. 
The depletion is largest for the most \mr stars.
The detailed \alp \abu distributions for these 2 very
different external \gals (one a dSph and the other a Magellanic irregular)
are quite      similar.
In addition, one can draw a single line in the Mg
and Ca plots that does a good job of fitting both \gals with no
evidence for kinks. The slope of this line is very similar to that shown
in our \Glx for the transition from the pure halo (at low \feh) to pure
disk (at high \feh), but is offset to lower \mets in the dwarfs.
However, there are some important differences between the behavior of these two
\gals in these diagrams. With the exception of a single star in Mg, the largest
enhancements occur in the LMC. In Mg and Si, Scl has several stars that are
more depleted than any LMC stars. Thus, the LMC stars are in general less
distinct from their halo counterparts than Scl stars.
We will return to the \alp elements in more detail in Section 5.

Moving from the \alp elements, the next element considered as a comparison
species is sodium.  This element is selected as it is well-represented
in the dwarf galaxy results to date and its behavior looks different
than in the Milky Way in certain respects (S03).
Sodium provides somewhat different insights into chemical evolution
than oxygen and the other \alp elements; although it is primarily a product of SNe II in most stellar
populations, its yield is metallicity dependent.  The main source of
Na is carbon-burning, however it is also both produced and destroyed by proton
captures.  The final yield from SNe II then depends on the p/n ratio,
which is itself a function of metallicity, decreasing as the overall
metallicity increases.  In Figure 9 Na is combined with O and [Na/Fe]
is plotted versus [O/Fe].  Galactic results are shown as the small blue
symbols, with the disk represented by asterisks, the thick disk by
open squares, and the halo by open circles.  The dwarf galaxies are
the red and magenta solid points. In this diagram, the dwarf galaxies
clearly segregate, on average, from the Milky Way results.  The
solid green lines are schematic representations of what sort of
chemical evolution would occur in simple, extreme examples.  Starting
at an elevated value of [O/Fe] and lower [Na/Fe] (as would be expected
in a metal-poor environment that had been dominated chemically by
SNe II ejecta) the vertical line is what could be expected approximately
from pure SNe II ejecta being instaneously recycled into new massive
stars; the oxygen and iron yields are not terribly metallicity sensitive,
while the sodium yield increases as the metallicity increases, and p/n
decreases.  By-and-large the Milky Way halo stars follow such a pattern.
The solid line leading towards equally decreasing values of [O/Fe] and
[Na/Fe] would result from the addition of pure Fe and is meant to
mimic approximately the contribution from pure SNe Ia ejecta.  Within this
diagram, the differences between the dwarf galaxies and the Milky Way
populations can again be understood as being dominated by a population
of low-metallicity SNe II (where such stars would form from the much slower
increase in metallicity due to inefficient star formation) and a higher
proportion of SNe Ia.  Very few Galactic halo stars overlap the dwarf galaxy
results in Figure 9.

The above illustrated abundance ratios demonstrate that different
galactic environments produce distinct behaviors in how [x/Fe] varies from
one element to another.  Using iron as a fiducial element,
however, can complicate certain comparisons as Fe has
substantial contributions from both core-collapse supernovae
of Type II and the longer-lived binary supernovae of Type Ia.
The differences, as discussed above and below, quite probably are dominated
by differing star formation rates, or the efficiency at which
a galaxy can convert its gaseous reservoir into stars, as well as the 
presence and strength of galactic winds.  It can be
useful to also investigate abundance ratios that more nearly isolate
elemental yields from a single source.  Such a comparison is attempted in
Figure 10 where oxygen, not iron, is used as the main metallicity
indicator.  In this figure values of [Na/O] are plotted versus
[O/H].  In the top panel only field stars from the Milky Way
populations and the dwarf galaxies are shown.  In this case,
unlike Figure 9 (where dwarf galaxies segregate clearly from
the Milky Way populations), the dwarf galaxy points fall within the 
Galactic halo stars.
Predicted model yields (e.g., Woosley \&
Weaver 1995) produce decreasing [Na/O] values as [O/H] (or overall
stellar metallicity) decreases, as is seen in the top panel of
Figure 10.  This demonstrates that in the field star populations of
both the Milky Way and the dwarf galaxies, Na and O production are
dominated by SNe II. In globular \cls, the Na/O ratio varies wildly due
to proton captures that reduce O and enhance Na. The location of these
processes remains uncertain, but possibly occurs in a previous generation
of massive AGB stars (Ventura \& D'Antona 2005). 

The bottom panel of Figure 10 plots the same field-star points as in
the top panel, but also includes results for two well-studied globular
clusters: M13 and M4 (with M13 being the lower metallicity cluster).
The abundances for the M13 stars were taken
from the paper by Kraft et al. (1997) and for M4 from
Ivans et al. (1997).
In the case of the globular clusters the well-established Na-O anti-correlation
stands out, with a larger Na-O abundance variation in M13 than in M4.
These abundance variations have been established even in unevolved main-sequence
stars in globular clusters and point to some type of chemical evolution
which occurs in the early environment of these clusters.  The abundance
variations also include F, Al, and Mg and point to a nucleosynthetic site
driven by proton captures at temperatures of roughly 5x10$^{7}K$
(e.g. Kraft et al. 1997).  It
is clear here, however, that globular cluster-like chemical evolution
in stellar populations can be isolated by using only Na and O.

The elements heavier than iron that are produced primarily by neutron
captures, either the r- or s-process, can be adequately represented
by two key elements, barium and europium.  Barium is an s-process
product, with about 85\% of its solar system abundance due to the
s-process, while europium is an excellent proxy for measuring
r-process contributions as some 97\% of solar system
Eu nuclei arose from r-process nucleosynthesis (Burris et al. 2000).
Most s-process elements are created as a by-product of neutron
captures that are driven by thermal pulses in AGB stars, while
the r-process neutron captures occur during SNe II (for a review
of the neutron-capture elements and their relation to stellar
evolution, see Wallerstein et al. 1997).  The abundance ratio of
Ba/Eu is thus an approximate measure of the relative importance of
AGB star chemical contributions, relative to SNe II.

Figure 11 summarizes results for [Ba/Eu] as a function of [Fe/H]
for various stellar populations from the Milky Way, a sample of
dwarf spheroidals, and the captured system $\omega$ Cen.  The long
dashed horizontal lines represent values for [Ba/Eu] for pure
r-process (bottom line) and pure s-process barium and europium
abundances as set by the solar system r- and s-process fractions.

Low-metallicity Milky Way stars exhibit subsolar
[Ba/Eu] values, closer to the pure r-process mixture, and this is
indicative of higher fractions of SNe II material characterizing
the old Galactic halo population.  A significant fraction of
Galactic halo stars show only r-process Ba/Eu ratios in the
interval of [Fe/H]$\sim$ -3 to -2.  As [Fe/H] increases, there
is a gradual increase in [Ba/Eu] up to [Fe/H]$\sim$-1.0 followed
by a steep increase as [Fe/H] approaches the solar value.  This
increase in [Ba/Eu] reflects the increasing contributions from
AGB stars as chemical evolution proceeds in the Milky Way.

The dwarf galaxy stars follow a different trend from the Galactic
members.  The extreme system is $\omega$ Cen where [Ba/Eu]
increases by a factor of 20-25 as [Fe/H] increases from -2.0 to
-1.5.  The more metal-rich stars in $\omega$ Cen exhibit a
heavy-element abundance distribution that is dominated by a
pure s-process origin, i.e., chemical evolution dominated by
AGB stars.  The dwarf spheroidal systems shown in Figure 11
follow a trend that is in-between $\omega$ Cen and the Milky
Way, indicating that AGB stars are more important in the
chemical evolution of these small galaxies, but not to the
extreme found in $\omega$ Cen.

\section{Discussion}  

\subsection{The Problem}

We return to a more detailed discussion of the \alp elements, which are
particularly important both observationally and theoretically. For the 
purposes of this discussion,
we will include O as well as Mg,
Si, Ca and Ti, since their nucleosynthetic origins and \abu
behaviors are generally similar (although differences may exist in detail
-- e.g. Shetrone 2004 -- but uncovering clear trends
 will require another level of
observational complexity)
and in order to compare with
previous work, e.g. Nissen and Schuster (1997 - hereafter NS97) and Gratton et al. (2003).
We will refer here
to  the ``[\alp/Fe]'' abundance or simply the \alp \abu 
as the mean of [O/Fe], [Mg/Fe], [Si/Fe], [Ca/Fe]
and [Ti/Fe] (or whichever of these is available if not all are determined, where
we require a minimum of 3 elements).

An intriguing trend for dSph stars to have lower \alp abundances than in the halo was
first noted by Shetrone et al. (2001 - hereafter S01) and given more weight by the observations of additional
dSphs by S03. In that work and in Tolstoy et al. (2003 - hereafter T03),
they combined their data for giants in 
seven dSphs  including their sample of 5 Scl stars and found
that over the \met interval from $\sim -1$ to --3, the dSph \alp abundances 
are typically 0.1 -- 0.3 dex lower than in the \Glx at the same \met and use
this as a strong argument against the standard Searle-Zinn scenario for the
formation of the \Gal halo from dSph-like objects. 
G05 added an additional 4 Scl stars which confirmed these results.

We can  now
examine this issue in much more detail and with larger samples by adding 
the new data on abundances in giants in the Sgr
dSph from Smecker-Hane
and McWilliam (2002), Bonifacio et al. (2004), and Monaco et al. (2005);
in Ursa Minor (Sadakane et al. 2004),
and the initial results for the first FLAMES dSph study (Tolstoy
2005 - hereafter T05), who derives \alp \abus for almost 100 stars in Scl.
These latter are still preliminary so we here only refer to her Figure 4.
We also include the
first results from similar
observations of a few stars each in several Local Group dwarf irregular 
galaxies, including NGC 6822 (Venn et al. 2001), WLM (Venn et al. 2003),
Sextans A (Kaufer et al. 2004)  and IC 1613 (Tautvaiseine et al. 2007)    
as well as LMC cluster stars (J06, P06).
We also  average the O, Mg, Si, Ca and Ti abundances,
which yields              a single parameter describing \alp \abus
and  has smaller internal errors, allowing one to see any possible
trends more clearly. Finally, we can utilize new data from 
Fulbright (2002),    Stephens and Boesgaard (2002), Gratton et al.
(2003),
Ivans et al. (2003), Cayrel et al. (2004) and  Jonsell et al. (2005)
to improve on the sample of Galactic comparison stars.

In Figure 12 we show the full current dSph, dIrr, LMC and halo datasets.
Green symbols denote the various \Gal samples, described below.
We include the Scl stars from S03 and G05 as the blue asterisks.
%and the Scl FLAMES results from Tolstoy (2005) as the small blue
%asterisks. 
The blue stars are the Fornax dSph and the blue filled circles
are stars in other dSphs studied by S01 and S03.
Red filled circles are the Sgr sample of Smecker-Hane and
McWilliam (2002) and the red stars are the Sgr samples of
Bonifacio et al. (2004),    Monaco et al. (2005) and Sbordone et al. (2006). The cyan triangles are 
from the LMC clusters studied by J06, the cyan squares from the LMC field 
stars of P06 and the yellow
triangles are from the various dIrr studies.
We have also searched the literature to include  high resolution studies
of halo stars, especially those that have attempted to investigate \abus of
stars that are the most likely to have been accreted, based on their kinematics.
We first
utilize the data of NS97.
They studied a
number of stars in the halo and thick disk and found an interesting group  of 
halo stars with
significantly lower [\alp/Fe] than other stars of the same \met. These ``low
\alp" stars also have different kinematics and NS97 suggested they are good
candidates for stars having been accreted from dwarf galaxies with a different
chemical evolution history than that of the general halo. 
We also add data from Gratton et al. (2003) 
who have essentially
repeated the NS97 analysis but with a significantly enlarged sample. They divide their
halo stars  into two kinematic samples: those with significant galactic
rotation which
they term the dissipative collapse component and those without rotation or with
counter-rotation, which they refer to as the accreted  component. They find a 
significantly lower \alp \abu at a given \feh for the second component 
relative to the first. 
Figure 12 includes the Gratton et al. (smoothed) mean for their large sample of dissipative collapse
component stars as the solid green curve, which is meant to represent the 
``normal" halo, their accreted  component stars as the
green stars and the NS97 low \alp stars as the small green filled circles.
Fulbright (2002 - hereafter
F02) studied a large number of ``\alp - poor" halo stars. F02 found that
the \alp \abus correlated well with kinematics such that the most
extreme \alp -poor stars had the largest \Gal rest-frame velocities, after
dividing his sample into 3 velocity bins, with $\sim$ 60 stars per bin.
The largest velocity stars should be
the closest known \Gal counterparts to the dSph stars. We show the mean
values for his three components as the large green filled circles, plotted
at their mean values, where the
point with the largest \alp \abu is the lowest velocity bin and the smallest
\alp \abu is for the largest velocity stars.
Stephens and Boesgaard (2002) derived detailed \abus for 56 halo stars
kinematically selected to be the most likely candidates for being accreted,
possessing either a very large apogalactic distance,  a very large maximum
distance above the plane, and/or a very large retrograde orbital velocity.
Their mean points for 3 different \met bins, with from 3 to 32 stars per bin,
are given as the large green open circles. Ivans et al. (2003) investigated
\abus in 3 stars known previously to have low \alp \abus - these are shown
as the green triangles. Cayrel et al. (2004) have begun a large-scale study
of \abus in very \mp stars - the First Stars project - and their initial results are
given as the green filled squares.
Jonsell et al.'s (2005) sample
of 43 halo stars are the green open squares. 

Let us first look at the Scl sample, as it is by far the largest available
to date for an external system, thanks especially to the FLAMES data.
If there are any differences between \gals, it is best to study them 
individually.
The scatter seen in the Scl stars, including the T05 plot,
is rather small except at the \mp end, where there are only a few
stars. Note that the S03 and G05 small sample of stars nicely delineates the
main features.
There are several striking features about the Scl sample: the first
is the clear trend of decreasing \alp abundance
%with \met for the stars more \mr than --1.5 (although one could also claim
with  \met, with a uniform, $\approx$
continuous decrease.
This is the first time that such a trend has been seen so clearly in a dSph.
A  decrease  also
occurs in the Galaxy but starting only at a \met  $>-1$.
The second striking feature is that all Scl stars except for a handful of the most \mp FLAMES giants
are depleted in their \alp \abu relative to the \Glx. 
At the \mp end the 2 samples merge but begin to diverge by \feh$\sim -2$
and the difference with the \Glx is  particularly striking for the
most \mr stars, which are some 0.5 dex lower in [\alp/Fe]
than their \Gal counterparts. Based on a large-scale Ca triplet study (Tolstoy
et al. 2001), we are confident that the full \met range of this galaxy is
represented. Unfortunately, then, we cannot say whether the \alp \abus
continue to drop at the highest \mets or are higher at the lowest \mets as there
are simply no, or very few, stars at these \mets in Scl.

Next we     look at the full ensemble of \gals depicted here.
It was already clear from the work of
S01, S03 and G05      that the dSphs have depleted
\alp abundances compared to the typical halo star
and the present dataset shows this even more clearly.  They also pointed out
the very interesting fact 
that the different dSphs display very similar behavior in this diagram,
despite their widely varied star formation histories. However, we can now
point to the very important exception of Sgr, which does display unique
behavior in this diagram.

Are there ANY stars in our halo that have abundances like those in dSphs?
Until now, there was little overlap in the \mets of dSph stars from the samples
of Shetrone and collaborators with those of the low \alp stars of NS97.
But the  G05 and Tolstoy et al. (2004) \mr Scl stars and the Sgr stars, as well as the addition of the Gratton et al. (2003)
sample, now remove this problem and show some 
fascinating trends. Starting at the most \mp end, there are still only 2
stars observed at high resolution
in dSphs with $\feh <-2.4$. These 2 stars have depleted
\alps, although note that there is at least 1 star in the Cayrel ``First Stars'
sample with similar \met and depletion.
For $-2.4 \ltsim \feh \ltsim -1.6$, the bulk of the dSphs stars are slightly
to significantly depleted, although a few Scl FLAMES stars appear similar 
to the halo. There is a very occasional halo star that falls amongst the bulk
of the dSph stars.
One such star         is  the extreme halo star
$BD+80^\circ 245$, which has \feh$=-1.86$
and $[\alpha/Fe]=-0.29$ (Carney et al. 1997, Ivans et al. 2003).
This places it near several of the most extreme low
\alp dSph stars (at slightly higher \mets). However, this star lies one or two orders of magnitude below
dSph stars in its Ba and Eu \abus, arguing against their common origin.
From $-1.6 \ltsim \feh \ltsim -1.0$, the situation is grim indeed - there is
essentially no overlap between the 2 samples, with the difference being very
substantial, with the very notable exception of Sgr. Sgr alone shows actually
quite good agreement (with a slight depletion)
with the halo for the 6 stars for which good \abus exist.
Clearly it would be extremely useful to observe more \mp stars in Sgr to see
if this trend continues. 
What is distinct about Sgr is of course that it is the                          most massive dSph known, with a mass at least several times greater than the    next most massive dSph, Fornax, and many times larger than the other dSphs.     
Finally, only Sgr has stars more \mr than $\feh=-1$
with the exception of a single Fornax giant which is very halo-like.
However, here we have a small problem -- there appears to be an offset of about 0.15 -- 0.2     dex between the
\alp \abus of Sgr stars at a given \met derived by Smecker-Hane and McWilliam
and those of the Italian group (Bonifacio et al. 2004, Monaco et al. 2005, Sbordone 
et al. 2006).
Note that this offset does NOT exist for the (small number) of \mp stars.
We are uncertain as to the origin of this offset (see Monaco et al. for more
details).
We will take the \abus at their face values.
The data of Smecker-Hane and McWilliam suggest there may be some overlap
with the most extreme low-\alp, high \met halo stars, while the Sgr 
stars measured by the Italians     continue the trend shown by the other
dSphs of low relative \alp \abus, although of course by the time we reach
solar \met in the \Glx, the mean \alp \abu is also solar and thus even the
Italian results show only a minor depletion of $\sim 0.15$ dex at this \met.
These results build on those derived by Venn et al. (2004) who carried out
a very detailed comparative study of the chemistry in different \Gal
components, separated by their kinematics, and dSphs.

Thus, it appears that only the most massive dSph has stars similar in \alp    \abu to \Gal stars, even the most extreme low \alp subset most likely to have
been accreted.
There appear to be very few stars in the less 
massive, more typical dSphs that overlap in this property with any stars in the \Glx.
This may be due to the likely possibility that Sgr was a dE galaxy rather than
a dSph (Monaco et al. 2005).
The difference in \alp \abu with the \Gal mean for all dSphs except Sgr
is  LARGER at higher \met, i.e. \feh
$\sim -1$, than at the lower \mets previously well
explored by S01 and S03. This 
is contrary to the suggestion by S03 that, although the \mp halo could not be
made up of stars like those seen in dSphs, up to 1/2 of the \mr halo could 
have originated in such objects. Our analysis
allows us to demonstrate that it {\bf appears 
very unlikely that a significant fraction of the \mr halo could have come
from disrupted dSphs of low mass similar to those studied herein}.
 However, it does appear possible
from this diagram that at least some of the \mp halo  may have come from
typical dSphs, and that a portion of the
intermediate \met and \mr halo
may have come from the accretion of very massive systems like Sgr.
%These two \pops also appear similar in their Na and hs \abus, which are the 
%only other elements we are able to compare with the available data.
However, McWilliam et al. (2003) and McWilliam and Smecker-Hane (2005b)
show that Sgr stars have a unique Mn/Fe  and Cu/Fe
signature and also that, at the \mr end, Sgr stars are depleted by $\sim 0.4$
dex in Na and Al with respect to their \Gal counterparts. 
Sbordone et al. (2006) confirm these findings and extend them to other 
elements including Sc, V, Co, Ni, and Zn. It is of paramount
importance to clarify the \abus in Sgr stars already measured and obtain new 
data, especially for its most metal-poor red giants.

We can help to fill in this critical \mr zone for comparison 
to the halo by utilizing recent
high resolution \abus derived for a small number of blue or red supergiants in each
of several Local Group dwarf irregular (dIrr) galaxies. It is often assumed 
that the dIrrs are the counterparts of dSphs which have managed to retain their
gas. Venn et al. (2001, 2003) and Kaufer et al. (2004) provide detailed \abus
for several stars in NGC 6822, WLM and Sex A, respectively, and we have included
new results on IC1613 (Tautvaiseine et al. 2007). These stars are shown
as the yellow triangles in Fig. 12.  
We also add the LMC cluster results from J06
as cyan triangles and the LMC field stars from P06 as cyan squares.
We see that the dIrr stars (granted only a
small number) appear to be rather uniform in their behavior. 
The dIrrs, as first noted by Venn et al. (2003), as well as the LMC stars (as 
expected from Fig 8),
follow the general trend of the dSph stars very well, although the 2 most \mp
dIrr stars, at $\feh\sim -1.2$ (in Sex A), are not as \alp depleted as the 
stars from the low mass dSphs,
lying between them and the Sgr stars of comparable \met, but still
depleted with respect to the bulk of the halo.

We are led to the rather amazing conclusion that {\bf the general \alp vs.
\feh pattern of 12 of the 13 \gals studied so far besides our own are similar to
each other and very different from the \MW!!} Sgr appears to be the only possible
exception.

These comparisons add renewed weight to the arguments first made by S01
and subsequently by F02, S03, T03 and G05
that the {\bf chemical compositions of stars
present now in typical low mass
dSphs are distinct from those in the \Gal halo} and disk.  Our analysis        
adds to this argument by enlarging the sample and 
independently confirming this trend, and extending
the \met regime to which this applies 
to higher \met; viz. we argue against the possibility suggested by
S03 that some 50\% of the \mr halo may have come from disrupted dSphs.
In addition, however, we suggest
that high mass dwarf systems like Sgr may potentially
be the source of some of the halo.

A comparison with globular clusters shows that there are a few
$\alpha$-poor systems such as Pal 12 (Cohen 2004) and Ruprecht 106
(Brown, Wallerstein, \& Zucker 1997).  However the strongly retrograde
globular NGC3201 does not show an $\alpha$-deficiency (Gonzalez \& Wallerstein
1998).  For $\omega$ Cen,
the abundance history is extremely complex (Norris \& Da Costa 1995).
For the globulars associated with the Sgr system, Terzan 7 has a
slightly lower [$\alpha$]/Fe] value at [Fe/H]= -0.6 than the Galactic
field stars.  Similarly, M54 (Brown, Wallerstein, \& Gonzalez 1999)
shows a mean [$\alpha$/Fe] value of +0.2, about half of the value for
field stars of similar metallicity, viz. [Fe/H]= -1.55.
In general, though, GC stars and halo field stars have very similar
chemical properties (except for the notorious Na -- O -- Al behavior in some GCs)
and therefore shared the same chemical evolution history, which was 
different from that of typical dSphs (Pritzl, Venn \& Irwin 2005).

The major implication of this of course is that {\bf the standard hierarchical
galaxy formation scenario and 
the Searle-Zinn paradigm for the formation of the outer \Gal halo via accretion
of ``fragments" composed of stars like those we see in
typical present-day dSphs is ruled out by the disjoint chemical signatures.
We refer to this as ``The Problem".}

Can we tell if a given star in the halo originated from a dSph?
As emphasized by S01 and S03 and further corroborated by this paper, dSph stars
have broadly similar \abu patterns which are generally quite
different from those found so far
in our \Glx.
The distinct \abu signature of dSph stars, as well as kinematic
differences expected for a disrupted system, suggest that it would be a rather
unambiguous task to determine if a given star or stellar structure may have 
originated from a dSph like those studied here. An obvious first starting 
point is to search for \mp stars ($\feh<-1$) with $[\alpha/Fe]<0.05$, as
also suggested by Font et al. (2006a,b). Sgr stars could still be missed by
this criterion. But
McWilliam  et al. (2003) and McWilliam and Smecker-Hane (2005b) 
suggest even Sgr stars could be distinguished by using their unique [Mn/Fe] 
and Cu/Fe signatures.
%It would be of 
%great interest, e.g., to obtain high resolution spectra of stars found
%in the newly discovered stellar ring surrounding the disk of our
%\Glx (Yanny et al. 2003).
%Bullock and Johnston (2004)  have carried out N-body simulations of the 
%accretion of low mass \gals to form the \Gal halo. They find that stars 
%associated with the halo may be quite different chemically from stars in
%surviving satellites due to distinct chemical evolutionary histories and suggest
%this may account for the differences we find.

%\subsection{Comparative Galactic Chemical Evolution}
\subsection{Explaining the Problem}

The uniformity of the chemical \abu patterns in dSphs (S01, S03, T03, this work)
suggest a fairly uniform
chemical evolution, despite the rather large range of star formation histories
(e.g. Mateo 1998).
Here we explore chemical evolution scenarios in dSphs
and how their
evolution appears to have differed from that experienced in our \Glx.
We would especially like to understand the reason for the distinct 
chemical \abus seen today between halo and dSph stars, in particular the \alp elements, as described in the previous subsection.
A great deal of progress has been made in very recent years which allows us to
now understand the origin of these \abu differences.

The canonical description of \Gal chemical enrichment dates back to Tinsley
(1979). She describes the \abu patterns expected due to the varying ratio 
of SNe II to SNe Ia during the lifetime of the \Glx. SNe II arise from massive stars
with main sequence lifetimes $< 10^8$ yr and have ejecta rich in the
\alp elements as well as Na and probably Eu.
SNe Ia come from mass transfer onto a white dwarf, which explode
after they pass the Chandrasekhar limit, typically at least a Gyr after 
formation. Their primary ejecta are  Fe-peak elements. Thus, in an initial low
\met environment
characterized by a standard IMF, the second
generation of stars will exhibit enhanced \alp \abus (compared to solar), 
with $[\alpha/Fe]\sim 0.4$, the standard SNe II ratio. As
the Fe \abu slowly builds up, the \alp \abus will remain enhanced at about this 
level until the lower mass SNe Ia begin to explode, $\gtsim 1$ Gyr after the first
episode of star formation.
Thereafter [\alp/Fe]
will begin to decrease. This produces a ``knee" in [\alp/Fe] vs. \feh at an \feh
that depends on the chemical evolution rate (basically the star formation rate
and yield). If the star formation rate (herafter SFR)
is rapid, the galaxy will have achieved a
relatively higher \feh within a given time
than would a system like a dSph experiencing a lower SFR or an early burst
followed by a long quiesence (e.g. Gilmore and Wyse 1991).

A complication arising in dSphs that is still to be solved is how such a dwarf
galaxy was able to retain the products of powerful SNe II ejecta
so that they could
be incorporated in the next generation of stars, as a \gal wind is expected
to be generated  which could efficiently blow out any remaining gas
(although see Marcolini
et al. 2006).
Clearly, the loss of processed stellar ejecta (via a galactic wind),
or infall of matter, can also influence the shape of the
[\alp/Fe]--[Fe/H] relation and also prevent star formation.
%the primary effect results
%from the star formation efficiency.

In the Milky Way, the SFR                 was relatively rapid and the \Glx
produced \alp-enhanced stars and 
enriched itself to \feh$\sim -1$ within the timescale for
the first SNe Ia to go
off, subsequently producing the knee seen in Figure 7. 
In Scl, the best observed dSph, we see a similar process - 
although it is not really clear what is the general behavior for \mp stars,
they are certainly \alp enhanced 
and may have a relatively shallow decrease in \alp with 
\met followed by an apparent
`knee' at higher \met and a subsequent steeper decline of [\alp/Fe].
The main difference with the halo is that 
the `knee' occurs at a much lower  \met than in the \Glx, $\sim -1.7$. 

The most straightforward interpretation of
the \feh value of the knee is  that Scl had a lower SFR                 than 
in the \Glx and only achieved an \feh of $\sim-1.7$ before
the onset of SNe Ia explosions.
It is well known qualitatively that dSphs must have a lower SFR (e.g. Gilmore
and Wyse 1991). 
Lanfranchi and Matteucci (2004 - hereafter LF04) have modelled these effects in detail
and find that {\bf the chemical differences between the dSphs and the halo are 
easily and simply understood 
as due to a combination of
a low star formation efficiency ($\sim$ SFR)
and  a high galactic wind efficiency in the former}. They carry out chemical evolution models
with a variety of parameters and try and match the observed \abus, star 
formation histories and age-\met relations for the Local Group dSphs. They
find that in every case except for Sgr, a very good match is derived when
the star formation efficiencies are very low and the wind efficiencies high. 
Sgr requires a substantially higher star formation efficiency.
They predict both a depressed \mp plateau as well as the \mp knee. The knee is
due both to the onset of SNe Ia as well as the occurence of the galactic wind
which blows   out much of the available remaining gas and halts further star
formation and subsequent SNe II.
For example, the O, Mg, Si and Ca \abus of Scl available to them - those of
S03 - are well fit by a star formation efficiency of 0.05 -- 0.5/Gyr and a 
high wind efficiency. We now add the G05 \alp data for Scl
and have combined the predictions for individual elements from the best fit
Scl model of LF04 into
a single \alp curve. The results are shown in Fig. 13. 
Again, we include in our discussion a comparison with the FLAMES results of T05.
The predicted
knee at $\feh=-1.6$
is very pronounced and fits the data reasonably well, especially when the 
FLAMES results are included. The smaller slope at lower \met
is not as clearly defined by the data, which show a larger spread in this
\met regime. The steep slope at higher \met fits                     the data,
although the data show a somewhat shallower decrease. The overall agreement
is reasonable and lends credence to the LF04 analysis.  We note that their
models also do quite well in fitting the sparser data for other dSphs.
However, note that the Scl data could be fit even better by a model producing
a constant decline in [\alp/Fe] with \feh.

The success of their model in explaining the \alp \abus prompted Lanfranchi
et al. (2006) to also apply it to the evolution of Ba and Eu in dSphs. Using
the same best fit parameters found for the \alp \abus, i.e. star formation
efficiency and \gal wind efficiency, they find similar
good fits to these important heavier elements in the dSphs with the best 
data available, including Scl. We reproduce here as Figure 14
their figure for Scl but
have included 2 additional stars that they did not which shows an excellent
fit to the data. They find the same kind of knee as for 
the \alp elements, at the same \met and for the same reasons. Again, the roles of SFR and \gal
winds are critical. 
Both elements are produced early on in
the lifetime of a galaxy in relatively constant amounts by SNe II, yielding the
horizontal line at low \met. However, when SNe Ia begin to kick in after about a Gyr
of evolution, a \gal wind is generated which stops further SF and hence SNe II.
Thus, no more Eu is produced but Ba slowly begins to rise as the contribution
from AGB stars becomes important. A similar rise is again seen in the \Glx
but at higher \met (see Fig 11) due to the higher SFR there. 

Thus, {\bf we are led to a simple explanation for the differences in \abu behavior
that we see between the halo and its nearest \gal neighbors: the halo
experienced much quicker chemical evolution due to its higher SFR, and the 
dSphs, at least the typical low mass ones, experienced strong \gal winds which
surpressed further massive star formation after SNe Ia began to explode.}
The latest models of Lanfranchi and
collaborators account quite well for the observed differences.
 
However, such models do have at least one glaring failure: their
predicted \met distribution does not mimic             observations.
Both the large samples derived from Ca triplet studies (Tolstoy et al. 2001)
as well as the detailed \feh values based on     FLAMES results
(Tolstoy et al. 2004) for Scl show that the 
observed \met distribution is  essentially bimodal, with \mp and \mr components
with the dividing line  at $\feh=-1.7$. The \mp stars are more spatially
extended and are kinematically hotter than the \mr stars. On the other
hand, Lanfranchi \& Matteucci produce only a single population with a 
peak at $\feh=-1.7$ and a broad dispersion. The discovery of multiple \pops
in even these simplest stellar conglomerates adds a new wrinkle to
this whole area. Clearly, if there are 2 components, \abus need to be derived
separately for each one. However, note that Kawata et al. (2006) find that a
simple system with a continuous, albeit steep, \met gradient can lead to
observations which suggest the system is instead discontinuous and bimodal.
Clearly, more observations are required to investigate this possibility.
In any case, we are now able to theoretically account for the observed 
chemical differences between the halo and dSphs.

\subsection{Solving the Problem}

We are still left with the very
serious Problem - we cannot build the halo out of the obvious dwarf
\gal ``fragments", the objects that otherwise strongly favored cosmological theories
suggest should be the basic building blocks of \gal structure.
Can we solve this Problem?

A series of recent papers have made great strides in presenting a very 
viable solution to The Problem. These papers include Robertson et al. (2005 -
R05), Font et al. (2006a,b), and Bullock \& Johnston (2005).
We will mostly focus on the main arguments, presented in R05. They have 
combined chemical evolution models with cosmologically-motivated 
mass accretion histories for the putative \MW dark halo and its satellites.
They begin with representative examples of the type of dark matter haloes 
expected to host a destroyed ``stellar halo progenitor" dwarf, a surviving
dIrr and a surviving dSph. They then include star formation and chemical
evolution allowing for enrichment from SNe Ia and SNe II as well as stellar winds.
Their solution to The Problem relies on the $\Lambda$CDM prediction that
{\bf the majority of halo stars formed within a few ($\sim 5$), very massive
($\sim 5\times10^{10}M_\odot$) satellites accreted very early in the 
history of the \MW ($\sim 10$ Gyr ago). Being massive, these satellites,
before being assimilated, underwent rapid SF}. For the reasons developed above,
such objects should then show $\sim$ halo \abu patterns. They were subsequently
accreted by the growing \Glx, disrupted and their stars then became the halo
stars we see today. These massive dwarf systems were perhaps
more like dEs than dSphs.

What about the dSphs? They argue that {\bf the dSphs we see today are NOT
representative of the type of building block that was responsible for the 
bulk of the halo but instead are a biased \pop. 
They are 'survivors'} - objects which have lived most
of their life in isolation and are only now (in the last few Gyrs) being
slowly assimilated by the \Glx and are still pretty much intact. Thus, they
have undergone the processes noted by LF04 - low SFR and eventual onset
of a \gal wind. They are also generally low mass.
R05 find, as expected, that such \gals end up with depleted
\alp \abus, in good agreement with the results of LF04 and the observations.

Let us look in detail at the 3 different objects that R05 model.
The first is designed to represent a dIrr.
It starts with a total dark halo mass of $6\times10^{10}M_\odot$  
and ends with a final mass in stars of $4\times10^{8}M_\odot$. This is
similar to that of a typical dIrr like IC 1613.
This object formed 11.6 Gyr ago and was accreted
by the \Glx 3.1 Gyr ago. It thus had 8.5 Gyr of star formation and chemical
evolution in isolation, allowing the effects of SNe Ia and a \gal wind to
shape its \abus. This object ends up having a relatively low SFR,
forms few stars initially and the majority of its stars form at relatively
high \met ($\sim -0.6<\feh<-1$) with $[\alpha/Fe]\sim$ solar, similar to the
stars observed in present-day dIrrs.

Secondly, they study an object designed to mimic those expected to have
formed the bulk of the halo. Its  initial mass is identical to that of the
above dIrr. Being of the same initial mass, it formed at the same time 11.6
Gyr ago. However, this object has two main differences with the dIrr: first
(for reasons not specified by the authors), it forms stars initially
very quickly. Secondly, it is accreted by the proto \Glx also very early --
9 Gyr ago. Thus, this object had only 2.6 Gyr of independence, at a high
level of star formation. Under these conditions, they posit, or their models
predict, that SNe Ia and \gal winds have little effect on the chemical 
evolution, leading to stars with enhanced \alp \abus. Its gas enriches so
that at the time of accretion               a typical star
has    $\feh=-1.1$ and $[\alpha/Fe]\sim0.2$. This system, with a total
stellar mass very similar to that above, quickly disrupts
after it is accreted and subsequently forms typical halo stars (although 
this is a little low in $[\alpha/Fe]$, which is more like 0.35 at this \met in
the halo).

Their final object is modelled after a typical dSph. It has a dark halo mass
of $6\times10^{8}M_\odot$ and thus formed very early, 13.1 Gyr ago. Its
final stellar mass is $1\times10^{6}M_\odot$, an order of magnitude less
than an object like Scl. It was accreted some 5 Gyr ago, so had 8 Gyr of
isolation. As expected, such an object had a low SFR, experienced
\gal winds, injection of SNe Ia and AGB material and thus ends up with
the knee in the $[\alpha/Fe]$ vs. \feh diagram at low          \met
and depressed \alp \abus for \mr stars. The gas just prior to accretion 
has enriched to $\feh\sim -1.5$ and $[\alpha/Fe]\sim-0.1$. This is
similar to what is seen in the low mass dSphs. 

Thus, {\bf R05 can successfully account for the \abus that we see in the halo
and in dSphs and dIrrs and appear to have rescued the SZ/hierarchical
formation scenarios.} This is a very important step which certainly requires
further modelling and observational confirmation.

\section{Some Problems with Solving The Problem}

There are a couple of potential problems in the R05 analysis that we wish to
point out. Note that in order to account for the halo \abus they need to have
their halo progenitor form stars very quickly and efficiently after its
formation. The dIrr progenitor formed at the same time with the same mass
 but did not have this
initial burst of star formation. What is the reason for the different star
formation histories of these otherwise originally identical
objects? Possibly the halo progenitor had its
star formation triggered by a close encounter with the proto\Glx very early.

More problematical is the question of timescales. We know from the results
presented above that by an $\feh\sim -2$ a typical low mass dSph like Scl
starts showing The Problem. For stars more \mr than this, The Problem becomes
increasingly severe. How long does The Problem take to develop?  
The models of LF04 suggest VERY QUICKLY -- an object like Scl reaches this
\met in less than 0.5 Gyr after star formation commences. The observational
age-\met relation for Scl of Tolstoy et al. (2003) backs this up. Thus, any
halo building block with a star formation and chemical evolution history
roughly
similar to that of Scl (presumably all of the typical dSphs) will show The
Problem unless they are accreted extremely early in their history, i.e. 
within 0.5 Gyr of their formation, i.e. onset of star formation in them.
Of course, R05 argue that such \gals 
only comprise a small fraction of halo constituents. But we have shown that
this fraction must be {\it very} low given the essentially negligible overlap in
\abus between the halo and low mass dSphs for stars more \mr than $\sim -2$.
If even a relatively small number of typical dSphs were accreted into the
halo one might expect that this discrepancy would not be as pronounced as it is.

What about more massive dwarf systems, like Sgr? Observations indicate that 
there is reasonably
good agreement for stars up to $\feh\sim-1$, above which Sgr stars also start
to show The Problem. LF04 again find that Sgr should reach such a \met on a
very short, similar ($<0.5$ Gyr) timescale. So, although more massive dSphs
may be able to make halo-like stars up to a \met of about --1, The Problem
occurs for more \mr stars, and these stars will be present if the object
is accreted more than $\sim 0.5$ Gyr after its \form (although perhaps these
more \mr stars formed later than this,        as suggested by some CMD studies).
In their analysis, R05 found that the Problem did not occur in their stellar
halo progenitor because it `only' had 2.5 Gyr of evolution before being 
accreted and this apparently was not enough time for the effects of SNe Ia 
and \gal winds to be important according to their models.
However, real \gals (at least Scl) and the 
LF04 models suggest otherwise. Note that the usual assumption for the onset
of SNe Ia is $\sim 1$ Gyr so one would indeed expect their effects to be
significant after 2.5 Gyr.               

We are thus led to suggest that the R05 scenario, although very promising,
may require some fine tuning, especially with regards to timescales, in
order to prevent The Problem from arising. In particular, satellite \gals
must either be accreted MUCH earlier than postulated in R05 (within 
$\sim 0.5$ Gyr after their \form) or somehow SF must be prevented to occur
in them after their \form
until only shortly before they are accreted. 
However, note that all dSphs and dIrrs had at least some star \form at the very
earliest epochs. The R05 solution relies
on having only a few very massive halos accreted `very early' but in fact
these massive haloes must be accreted within a very short time after their
\form. 
{\bf If dSph-like objects were accreted to form much of the halo, they
must have been accreted very early, before the onset of SNe Ia}. Thus, we may be
left with a hybrid scenario where the \Glx may have collapsed as well as
accreted fragments a la Searle and Zinn but on an ELS-like timecale.

%Finally, as remarked by T03, it may still be possible to have \lq Sculptor'-ed
%the halo of our \Glx {\it if} the assimilation 
%ocurred very early, thereby allowing the majority of star formation to have
%occurred in the much larger potential, higher star formation efficiency and
%smaller effect of any galactic wind that are
%required to explain the abundance patterns seen in our Galaxy, but not found in
%dSphs. Note that the closer resemblance of the \alp \abus in Sgr derived by
%Smecker-Hane and McWilliam to those in the
%\Glx support this, although the Bonifacio et al.     data do not. But Scl shows 
%us that some of the \abu differences, viz. the depressed \alp values at the \mp
%end, are already in place before      the first SNe Ia's exploded, shortening
%the time of any potential mergers from the 
%``first few Gyr of structure formation" suggested in T03 to perhaps only a Gyr
%or less. Note that the Lanfranchi and Matteucci models predict that Scl took
%about a Gyr to reach the \met of $\sim -1.5$ where the knee occurs.
%{\it If less massive dSph-like objects were accreted to form much of the halo, they
%must have been accreted very early, before the onset of SNeIa}. Thus, we may be
%left with a hybrid scenario where the \Glx may have collapsed as well as 
%accreted fragments a la Searle and Zinn but on an ELS-like timecale.

\subsection{Other Considerations}

We close with a few brief remarks about other important factors or results
related to our general topic. First, virtually all of the field stars that have
been observed at high resolution to date to form our knowledge of the 
composition of the halo currently
inhabit only a very small region of the halo centered
on the Sun. 
How representative of the  full halo is this region? 
Much wider area surveys are required to reveal this and are being planned.
Lee \& Carney (2002)
suggested that, although the mean \alp \abus may be uniform, there may be
a gradient in [Si/Ti] in old halo GCs. However, Pritzl et al. (2005) compiled
high resolution \abus for 45 GCs and compared them to halo field stars and
those of dSphs. They found no evidence for any chemical gradients, in 
particular [Si/Ti], and indeed found that the detailed \abus of the GC and
field stars were very similar and that they then must have shared the same
evolutionary history. Cohen \& Melendez (2005) found that the detailed \abus
of stars in the distant outer halo GC NGC 7492 were the same as those in the
inner halo GCs M3 and M13, corroborating the Pritzl et al. result.

This raises the general question of the homogeneity of the halo. The early
results of McWilliam et al. (1995) confirmed the general  expectation that at very 
low \mets one should start seeing substantial differences in element ratios
due to stochastic events related to the mixing of one or only a few SNe of
different mass and hence nucleosynthetic output. However, as more and more 
data are acquired, of better resolution, S/N and sample size (e.g. the First
Stars results of Cayrel et al. 2004, Arnone et al. 2005 Beers \& Christlieb 2005), {\bf
the intrinsic
scatter in many elements, particularly the \alps, is approaching the
observational errors, which are reaching very low levels}.
Arnone et al. find that, with a very careful analysis
of very similar stars, that the cosmic scatter in Mg was $<0.06$ dex over a
sample of  23 main sequence halo stars ranging from $-3.4<\feh<-2.2$. This is
truly an amazing result. Standard \Gal chemical evolution models 
(e.g. Argast et al. 2002) predict
that there should be essentially no mixing and therefore substantial scatter 
at $\feh=-3$, with a total range in e.g. the Mg \abu at this \met of 1 dex
and a standard deviation of 0.4 dex, with the scatter decreasing with \met
and disappearing by $\feh=-2$. The real halo however is much more homogeneous than
this, with a cosmic scatter (at least in Mg) an order of magnitude lower than
expected. They find that {\bf the halo must have been very well mixed within $\sim 30$
Myr after its \form}, the estimated time for the \met to reach -3. This is an extremely short
timescale. Andersen et al. (2007) estimate that at least 30 SNe II must have
exploded to produce the homogeneity in the Mg \abus seen at \feh=-3.
Either the mixing time for the halo was much shorter than generally
assumed or the intrinsic variations in \abu ratios of SNe II of different mass
and energy are much smaller than generally assumed. 
Also note that Melendez \& Cohen (2007) have argued that the halo formed on a
timescale of $\sim$0.3 Gyr based on the absence of a contribution from 
intermediate mass AGB stars to halo Mg isotope ratios.
%GCs are OLD - but outer halo GCs are a bit younger in the mean???

This point raises an additional problem for SZ/hierarchical \form scenarios.
The halo now not only needed to have accreted but also to have mixed on a 
very short timescale. In fact, as pointed out by Gilmore \& Wyse (2004), 
such a remarkable homogeneity of the halo already seems to rule out a large
number of merger/accretion events. One expects that each fragment/satellite
galaxy should have its own distinct chemical history, albeit similar in many
details, and it seems incredibly contrived to imagine many or even a few
of these coming together on such a short time scale and giving rise to such
a uniform halo, unless of course they were mostly gaseous at the time of
accretion. 

Initial results of several ESO Large Programmes devoted to the study of 
kinematics and \abus in dSphs are now appearing and will certainly revolutionize
this field. One of the first has important implications for this Review: Helmi
et al. (2006) report on their Ca triplet \mets for several hundred stars in each
of four dSphs. They find NO stars with \feh$<-3$ and that the \mp tail of the 
dSph \met distribution is thus significantly different from that of the halo. However,
$\lambda$CDM chemical evolution models (Prantzos et al. 2006) generally predict
that the majority of low \met halo stars should come from low mass, dSph-like
progenitors. This adds an extra wrench into halo building block theories.

A further argument against the SZ scenario comes from comparing the Oosterhoff
types of RR Lyraes in the halo (field and GCs) and dSphs. While \Gal RR Lyraes
very nicely separate into two classes -- Oo type I and II, with mean periods
of 0.55 and 0.65 days  -- those in the dSphs generally lie inbetween these 2 types (Catelan 2006).
It appears virtually impossible to obtain the clear separation seen in the
halo from the concatenation of a variety of objects with generally intermediate
properties.

It seems most likely from the above considerations that if the SZ/hierarchical
\form models are correct, that the bulk of the halo came from the disruption
of only a small number of very massive building blocks very early in the
history of the Universe, as suggested by R05 but much earlier than they envisioned.
Not only should 
such massive fragments contain the chemistry most like that seen now in the
halo but they are also the ones that will host their own GCs. Of the present
day dSphs, only the two most massive -- Fornax and Sgr -- possess GCs. Since
it is often suggested that many of the halo GCs, especially those in the 
outer halo, were accreted, one requires rather massive dwarf \gals  to provide
these GCs, as Sgr is now doing. 
Interestingly, Fornax GCs have \abus very similar to typical halo GCs
(Letarte et al. 2006); however,
as seen in Section 3, the Sgr GCs show The Problem.
Thus, although the \Glx was not
`Sculptor-ed' (see G05), it may well have been `Fornax-ed' or possibly `Sagittarius'ed'.

Indeed, perhaps GCs themselves are the missing halo building blocks. They
certainly contain the right stellar \pops, \abus, etc., by definition.
 Many theorists  have
argued that a substantial fraction of halo field stars are the disrupted remnants
of former GCs. E.g. Kroupa \& Boily (2002) studied the dynamical evolution of
\cls and suggested that the halo field stars were the disrupted lower-mass \cls
of an initial \cl \pop and the remaining intact GCs form the massive end.

Finally, we note that the classical, tacit assumption that the \MW is a typical
spiral galaxy may indeed not be correct. Recent evidence is beginning to suggest
that the \MW may in fact be quite unrepresentative. For a given mass, the 
Galaxy lies \gtsim $1\sigma$ below the mean of comparable local spirals in 
terms of its angular momentum, disk radius and \met in its outer regions
(Hammer et al. 2007). They suggest that this is because the \MW has had an
exceptionally quiet merger history, while most spirals have undergone
significantly more mergers and that M31 is a much more typical spiral.
Clearly, the suitability of our own Galaxy as a representative prototype is of
utmost importance and needs to be clarified.
Also, we still have much to learn about the nature of our own halo. A very recent
paper (Carollo et al. 2007) presents the clearest evidence to date from studies
of individual stars that the halo is composed of two subcomponents, with
different spatial and metallicity distributions and kinematics, as first
suggested by Zinn (1983) from GC studies. There is certainly much more to
discover about our Milky Way, in which the current and future generations of
surveys will play a leading role. 

A quote by Shapley (1943), in announcing the discovery (Shapley 1938) of a new class
of objects called dSphs, is as relevant today, in the context of galaxy
formation ideas, as it was originally regarding the nature of galaxies:
``the discovery of dSphs (Scl and Fornax) is upsetting,
because it implies that our former knowledge and assumptions concerning
the average galaxy may need serious modification ... 
Two hazy patches on a 
photograph have put us in a fog." We hope this Review has shed a little
light on this fog and suggested some new avenues of research where the fog
persists or has thickened.

We would like to thank Marta Mottini for helping to prepare some of the
graphs. D.G. gratefully acknowledges support from the Chilean 
{\sl Centro de Astrof\'\i sica} FONDAP No. 15010003.  V.V.S. acknowledges
support from the NSF through grant AST06-46790.  We would like to
thank E. Tolstoy for sending us data in advance of publication,    G.
Lanfranchi for sending us detailed theoretical models and J. Johnson for 
sending her results in machine-readable form.
M. Shetrone, K. Venn, E. Tolstoy,
V. Hill  and
collaborators are warmly congratulated for pioneering the field of 
extragalactic stellar abundances. 
Any research involving Galactic GCs is aided by the excellent database of B.
Harris who deserves special recognition for this effort.
Finally, D.G. would like to thank M.E.
Barraza for her enduring support, patience and love.

\clearpage

%\begin{thebibliography}{}

\references

\noindent
Andersen, J. 2007, IAU Symp. 241, Stellar Populations as Building Blocks of
Galaxies, ed. A. Vazdekis \& R.F. Peletier, in press\\
%Arnett, W.,D. 1996, Supernovae and Nucleosynthesis, Princeton University Press,
%Princeton\\ 
%Arlandini, C., Kappeler, F., Wisshak, K., Gallino, R., Lugaro, M., Busso, M.,
%\& Stariero, O. 1999, ApJ, 525, 886\\
Argast, D., Samland, M., Thielemann, F.-K., \& Gerhard, O.E. 2002, A\&A, 388, 842\\
Arnone, E.; Ryan, S. G.; Argast, D.; Norris, J. E.; Beers, T. C. 2005, A\&A, 430,
507\\
%Baade, W. 1944, ApJ, 100, 137\\
%Baade, W., \& Swope, H. H. 1961, AJ, 66, 300\\
%Bard, A., Kock, A., \& Kock, M. 1991, A\&A, 248, 315\\
Beers, T.C., \& Christlieb, N. 2005 ARAA 43, 531\\
Belokurov, V.; Zucker, D. B.; Evans, N. W.; Gilmore, G.; Vidrih, S.; Bramich, D. M.; Newberg, H. J.; Wyse, R. F. G.; Irwin, M. J.; Fellhauer, M.; et al. 2006a,
ApJ, 642L, 137\\
Belokurov, V.; Zucker, D. B.; Evans, N. W.; Wilkinson, M. I.; Irwin, M. J.; Hodgkin, S.; Bramich, D. M.; Irwin, J. M.; Gilmore, G.; Willman, B.; et al. 2006b,
ApJ, 647L, 111\\
%Bessell, M. S., \& Brett, J. M. 1988, PASP, 96, 247\\
%Bessell, M. S., Castelli, F., \& Plez, B. 1998, A\&A, 333, 231\\ 
%Bonifacio, P., et al.  2000, A\&A, 359, 663\\
%Bonifacio, P., \& Caffau, E. 2003, A\&A, 399, 1183\\
Bonifacio, P. Sbordone, L., Marconi, G., Pasquini, L. \& Hill, V. 2004, A\&A,
414, 503\\
%Bouchard, A., Carignan, C., \& Mashchenko, S. 2003, AJ, 126, 1295\\
Brown, J., A., Wallerstein, G. \& Zucker, D. 1997, AJ 114,180\\
Brown, J., A., Wallerstein, G., \& Gonzalez, G. 1999 AJ 118,1245\\ 
Bullock, J.S., \& Johnston, K.V. 2005, ApJ, 635, 931\\
%in Satellites and Tidal Streams, PASP Conf. Series 327, 801. Ed by F.Prada, D. 
%Martinez Delgado, \& T. J. Mahoney\\
Burris, D.L., Pilachowski, C.A., Armandroff, T.E., Sneden, C., Cowan, J.J., 
Roe, H. 2000, ApJ, 544, 302\\
%Carignan, C., Beaulieu, S., Cote, S., Demers, S., \& Mateo, M. 1998, AJ, 116,
%1690\\
Carney, B.W., Wright, J.S., Sneden, C., Laird, J.B., Aguilar, L.A., \& Latham,
D.W. 1997, AJ, 114, 363\\
Carollo, D., Beers, T.C. et al. 2007, Nature, in press (astro-ph/0706.3005)\\
Carraro, G., Zinn, R., \& Moni Bidin, C. 2007, A\&A, 466, 181\\
Carretta, E.; Gratton, R. G. 1997, A\&AS, 121, 95\\
%Carretta, E., Gratton, R., Cohen, J. G., Beers, T. C., Christlieb, N. 2002, AJ,
%124, 481\\
%Carpenter, J. M. 2001, AJ, 121, 2851\\
%CC_Sept5
Catelan, M.; Bellazzini, M.; Landsman, W. B.; Ferraro, F. R.; Fusi Pecci, F.; 
Galleti, S. 2001, AJ, 122, 3171\\
Catelan, M. 2006, RevMexAstAst, 26, 93    \\
Cayrel, R., Depagne, E., Spite, M., Hill, V., Spite, F., Francois, P.,
Plez, B., Beers, T., Primas, F., Andersen, J., Barbuy, B., Bonifacio, P.,
Molaro, P., Nordstrom, B. 2004, A\&A, 416, 1117 \\
Chaboyer, B.; Demarque, P.; Sarajedini, A. 1996, ApJ, 459, 558\\
%Charbonnel, C., Brown, J.A., Wallerstein, G., 1998, A\&A, 332, 204 \\
%Charbonnel, C., do Nascimento, J., 1998, A\&A, 336, 915 \\
%Charbonnel, C., \& Palacios, A., 2003, Proceedings of the IAU Symposium 215 
%on {\sl Stellar Rotation}, Eds A.Maeder \& P.Eenens \\
%Charbonnel, C. \& Weiss, A. in IAU General Assembly, Joint discussion 4, ed F. D'Antona \& R. Gratton\\
%CC_Sept5
Cohen, J.G. 2004, AJ, 127, 1545\\
Cohen J.G. \& Melendez, J. 2005, AJ, 129, 1607\\
%Cole, A.A., Smecker-Hane, T.A., Gallagher, J.S. 2000, AJ, 120, 1808\\
%Cunha, K., Smith, V.V., Suntzeff, N., B., Norris, J.E., Da Costa, G.,S., \&
%Plez, B. 2002 AJ 124,379\\
%Da Costa, G.S. 1984, ApJ, 285, 483\\
%Da Costa, G. S. 1988 in IAU Symp. 126, Harlow Shapley Symposium on 
%Globular Cluster Systems 
%in Galaxies, ed.       J. E. Grindlay \&  A. G. Davis Phillip (Reidel, 
%Dordrecht), p. 191\\
Da Costa, G. S. \& Armandroff, T. E. 1995, AJ 109, 2533\\
De Angeli, F., Piotto, G., Cassisi, S., Busso, G., Recio-Blanco, A.,
Salaris. M., Aparicio, A., Rosenberg, A. 2005, AJ, 130, 116\\
%Demers, S., \& Battinelli, P. 1998, AJ, 115, 154\\
%Dennisenkov, P.A., \& Weiss, A. 2001, ApJ, 559, L115\\
Dinescu, D. I., Girard, T. M., van Altena, W. F. 1999, AJ 117, 1792\\
Dinescu, D. I., Majewski, S. R., Girard, T. M., \& 
Cudworth, K. M., 2000, AJ, 120, 1892\\
Dinescu, D. I., Majewski, S. R., Girard, T. M., Cudworth, K. M., 2001, AJ 122, 19
16\\
Dinescu, D. I., Keeney, B. A., Majweski, S. R., Girard, T. M. 2004,
AJ 128, 687\\
Dinescu, D. I., Girard, T. M., Van Altena, W. F., Lopez, C. E. 2005, ApJ 618, L28
\\
%Dolphin, A.E. 2002, MNRAS, 332, 91\\
Edvardsson, B., Andersen, J., Gustafsson, B., Lambert, D. L., Nissen, P. E.,
Tomkin, J. 1993, A\&A, 275, 101 \\
Eggen, O.J., Lynden-Bell, D., \& Sandage, A. R. 1962, ApJ, 136, 748 (ELS)\\
Freeman, K., \& Bland-Hawthorn, J. 2002, ARAA, 40, 487   \\
Font, A., Johnston, K.V., Bullock, J.S., \& Robertson, B.E. 2006a, ApJ, 638, 585\\
Font, A., Johnston, K.V., Bullock, J.S., \& Robertson, B.E. 2006b, ApJ, 646, 886\\
Fulbright, J. P. 2000, AJ, 120, 1841 \\
Fulbright, J.P. 2002, AJ, 123, 404 (F02)\\
Fulbright, J. P. \& Johnson, J. A. 2003, ApJ, 595, 1154 \\
Geisler, D., Smith, V. V., Wallerstein, G., Gonzalez, G., Charbonnel, C.
2005, AJ, 129, 1428 (G05) \\
%CC_Sept5
%Geha, M., Guhathakurta, P., \& van der Marel, R.P. 2003, AJ, 126, 1794\\
Gilmore, G., \& Wyse, R.F.G. 1991, ApJL, 367, 55\\
Gilmore, G., \& Wyse, R.F.G. 2004, astro/ph-0411714\\
%Gilroy, K.K., 1989, ApJ, 347, 835 \\
%Gilroy, K.K., Brown, J.A., 1991, 371, 578 \\
Gonzalez, G., \& Wallerstein, G. 1998, AJ 116, 765\\
%Gratton, R.G., Sneden, C., Carretta, E., Bragaglia, A., 2000, A\&A, 354, 169 \\
%CC_Sept5
Gratton, R.G., Carretta, E., Caludi, R.,   Lucatello, S.,  \& 
Barbieri, M. 2003, A\&A, 404, 187 \\
%Gratton, R.G., Carretta, E., Desidera, S., Lucatello, S., Mazzei, P., \& 
%Barbieri, M. 2003b, A\&A, 406, 131\\
%Grebel, E.K., 1999, in IAU Symp. 192, The Stellar Content of Local Group
%Galaxies, ed. P. Whitelock \& R. Cannon (San Francisco: ASP), 17\\
%Grebel, E. K., Roberts, W. J., \& Van de Rydt, F. 1994, Third ESO/CTIO 
%Workshop, La Serena, Chile\\
Grebel, E.K., Gallagher, J., \& Harbeck, D. 2003, AJ, 125, 1926\\
%Grevesse, N., \&   Sauval, A.J. 1999, A\&A, 347, 348\\
%Gustaffson, B., Bell, R.A., Ericksson, K., \& Nordlund, A., 1975, A\&A, 42, 407\\
Harbeck, D., Grebel, E.K., Holtzman, J., Guhathakurta, P., Brandner, W., 
Dolphin, A., Geisler, D., Sarajedini, A., Hurley-Keller, D., Mateo, M. 2001, AJ,
122, 3092\\
Hammer, F., Puech, M., Chemin, L., Flores, H., \& Lehnert, M.D. 2007, ApJ, in 
press\\
Harris, W.E. 2003, Galactic Globular Cluster database\\
Helmi, Amina; Irwin, M. J.; Tolstoy, E.; Battaglia, G.; Hill, V.; Jablonka, P.; Venn, K.; Shetrone, M.; Letarte, B.; Arimoto, N.; et al. 2006, ApJ, 651L, 121\\
%Hilker, M., \& Richtler T. 2000, A\&A, 362, 895\\
%CC_Sept5 
%Hill, V., Barbuy, B., Spite, M., 1997, A\&A, 323, 461 \\
Hill, V., Francois, P., Spite, M., Primas, F., Spite, F. 2000, A\&A, 364,
L19 \\
%CC_Sept5 
%Holweger, H., Bard, A., Kock, M., Kock, A. 1991 A\&A 249, 545\\
Hoyle, F.; \& Schwarzschild, M. 1955, ApJS, 2, 1\\
%Hughes, J., \& Wallerstein, G. 2000, AJ, 119, 1225\\
%Hurley-Keller, D., Mateo, M., Grebel, E. 1999, ApJL, 523, 25\\
Ibata, R.A., Gilmore, G., \& Irwin, M.J. 1994, Nature, 370, 194\\
Ivans, I.,I., Sneden, C., Kraft, R.,P., Suntzeff, N.,B., Smith, V.,V., Langer, E. \& Fulbright, J.P. 1999 AJ 118,1273\\  
%Ivans, I.,I., Kraft, R.,P., Sneden,C., Rich, R., Shetrone, M. 2001 AJ 122,1438\\
Ivans, I.I.; Sneden, C.; James, C. R.; Preston, G.W.; Fulbright, J.P.; Hoflich, P.A.; Carney, B.W.; Wheeler, J.C. 2003, ApJ, 592, 906\\
Johnson, J. A. 2002, ApJS, 139, 219 \\
Johnson, J. A., Ivans, I.I., Stetson, P.B. 2006, ApJ, 640, 801 (J06)\\
Jonsell, K.; Edvardsson, B.; Gustafsson, B.; Magain, P.; Nissen, P. E.; Asplund, M.
2005, A\&A, 440, 321\\
Kaufer, A., Venn, K.A., Tolstoy, E., Pinte, C., \& Kudritzki, R.P. 2004, AJ,
127, 2723\\                    
Kawata, D., Arimoto, N., Cen, R., Gibson, B.K. 2006, ApJ, 641, 785\\
%Keller, Pilachowski and Sneden 2001, AJ 122, 2560???\\
Kleyna, J.T.; Wilkinson, M.I.; Evans, N.W.; Gilmore, G. 2005, ApJ, 630L, 141\\
Klypin, A.A., Kravtsov, A.V., Valenzuela, O., \& Prada, F. 1999, ApJ, 522, 82\\
Korn, A. J., Keller, S. C., Kaufer, A., Langer, N., Przybilla, N., Stahl, O.,
Wolf, B. 2002, A\&A, 385, 143 \\
Kraft, R. P., Sneden, C., Smith, G. H., Shetrone, M. D.,
  Langer, G. E., Pilachowski, C. A. 1997, AJ, 113, 279\\
Kroupa, P.; Boily, C. M. 2002, MNRAS, 336, 1188\\
Lanfranchi, G.A., \& Matteucci, F. 2004, MNRAS, 351, 1338 (LF04)\\              
Lanfranchi, G.A., Matteucci, F. Cescutti, G. 2006, MNRAS, 365, 477\\        
Layden, A.C. \& Sarajedini, A. 2000, AJ, 119, 1760\\
Lee, H-c, Lee, Y-W., Gibson, B. K. 2002, AJ 124, 2664\\
Lee, J-W. \& Carney, B.W. 2002, AJ, 124, 1511\\
Lee, Y-W., Demarque, P., Zinn, R. 1994, ApJ, 432, 248\\
Letarte, B.; Hill, V.; Jablonka, P.; Tolstoy, E.; Francois, P.; Meylan, G. 2006,
A\&A, 453, 547\\
%Lloyd Evans, D. 1983, MNRAS 204, 985\\
%Majewski, S.R., Siegel, M.H., Patterson, R.J., Rood, R.T. 1999, ApJL, 520, 33\\
Marcolini, A., D'Ercole, A., Brighenti, F., Recchi, S. 2006, MNRAS, 371, 643\\
%Martin, Fuhr, \& Wiese 1988???\\
Mateo, M. 1998, ARAA, 36, 435\\
%McWilliam, A. 1990, ApJS, 74, 1075\\
McWilliam, A., Preston, G. W., Sneden, C., Searle, L. 1995, AJ, 109, 1428 \\
%McWilliam, A. 1997, ARAA, 35, 503\\
McWilliam, A., Rich, R.M., Smecker-Hane, T.A., 2003, ApJL, 592, 21\\
McWilliam, A. \& Smecker-Hane, T. 2005a, ASP Conf. Ser. 336, p. 221\\
McWilliam, A. \& Smecker-Hane, T. 2005b, ApJ, 622, L29             \\
Melendez, J., \& Cohen, J.G. 2007, astro-ph/0702655\\
Minniti, D. 1995, AJ, 109, 1663\\
%Mishenina, T.,V., Kovtyukh, V.,V., Soubiran, C., Travaglio, C., \& Busso, M., 
%2002 A\&A 396,189\\
Monaco, L.; Bellazzini, M.; Bonifacio, P.; Ferraro, F. R.; Marconi, G.; Pancino, E.; Sbordone, L.; Zaggia, S.
2005, A\&A, 441, 141\\
%Monkiewicz, J., et al. 1999, PASP, 111, 1392\\
Navarro, J.F.; Frenk, C.S.; White, S.D. M. 1997, ApJ, 490, 493\\
Nissen, P.E., \& Schuster, W.J. 1997, A\&A, 326, 751 (NS97)\\
Norris, J.E. \& Da Costa, G.S. 1995 ApJ 447, 680\\
%Norris, J.E., Freeman, K.C., \& Mighell, K.J. 1996, ApJ, 462, 241\\
Odenkirchen, M.; Grebel, E.K.; Dehnen, W.; Rix, H-W.; Yanny, B.; Newberg, H.J.; Rockosi, C.M.; Martínez-Delgado, D.; Brinkmann, J.; Pier, J.R.
2003, AJ, 126, 2385\\
%O'Brian et al. (1991???\\
Pagel, B.E.J., \& Tautvaisiene, G. 1995, MNRAS, 276,
505\\
Piatek, S., Pryor, C., Olszewski, E. W., Harris, H. C., Mateo, M.,
Minniti, D., Monet, D. G., Morrison, H., Tinney, C. G. 2002, AJ 124, 3198\\
Pompeia, L., Hill, V., Spite, M., Cole, A., Primas, F., Romaniello, M., 
Pasquini, L., Cioni, M-R., Smecker-Hane, T. 2006, A\&A, in press (P06)\\
Prantzos, N. 2006, astro-ph/0611476\\
Pritzl, B.J., Venn, K.A. \& Irwin, M. 2005, AJ, 130, 2140\\
Prochaska, J. X., Naumov, S. O., Carney, B. W., McWilliam, A., Wolfe, A.
2000, AJ, 120, 2513 \\
%Ramirez, S.V., Cohen, J.G., Buss, J., Briley, M.M. 2001, AJ, 122, 1429\\
Reddy, B. E., Tomkin, J., Lambert, D. L., \& Allende Prieto, C. 2003,
MNRAS, 340, 304 \\  
Reddy, B. E., Lambert, D. L., \& Allende Prieto, C. 2006, MNRAS, 367, 1329\\
Robertson, B.; Bullock, J.S.; Font, A.S.; Johnston, K.V.; Hernquist, L. 2005, 
ApJ, 632, 872 (R05) \\
%Renzini, A. 1980, Mem. S.A.It., 51, 749\\
%Russell, S.C., Bessell, M.S., \& Dopita, M.A. 1988, 
%Galactic and Extragalactic Star Formation, Proceeings of a NATO Advanced Study Institute, (Dordrecht:Kluwer),
%edited by Ralph E. Pudritz and Michel Fich. NATO Advanced Science Institutes Series C, Volume 232, p.601\\
%Ryan, S.G., Norris, J.E., \& Beers, T.C. 1996, ApJ, 471, 254\\
Sadakane, K.; Arimoto, N.; Ikuta, C.; Aoki, W.; Jablonka, P.; Tajitsu, A. 2004,
PASJ, 56, 1041\\
Sandage, A.R. 1953, AJ, 58, 61\\
Sandage, A., \& Wildey, R. 1967, ApJ 150,469\\
Sarajedini, A. \& Layden, A.C. 1995, AJ, 109, 1086\\
Sbordone, L., Bonifacio, P., Buonanno, R., Marconi, G., Monaco, L. \& Zaggia, S.
2007, A\&A, 465, 815  \\
%Schweitzer, A. E., Cudworth, K. M., Majewski, S. R., \& Suntzeff, N. B. 
%1995, AJ, 110, 2747\\
Searle, L., \& Zinn, R. 1978, ApJ, 225, 357 (SZ)\\
Shapley, H. 1938, Nature, 142, 715\\
Shapley, H. 1943, Galaxies, Philadelphia: Blakiston\\
%CC_Sept5
Shetrone, M. 2003, ApJL, 585, 45 \\
Shetrone, M. 2004, 
Origin and Evolution of the Elements, from the Carnegie Observatories Centennial Symposia. Published by Cambridge University Press, 
Carnegie Observatories Astrophysics Series. Edited by A. McWilliam and M. Rauch, p. 218\\
%CC_Sept5
Shetrone, M. D., Bolte, M., \& Stetson, P. B. 1998, AJ, 115, 1888\\
Shetrone, M., D., C\^ot\'e, P., \& Sargent, W., L., W. 2001, ApJ, 548, 592 (S01)
\\
Shetrone, M., Venn, K., Tolstoy, E., Primas, F., Hill, V., \& Kaufer, A. 2003,
AJ, 125, 684 (S03) \\ 
Simon, J.D. \& Geha, M. 2007, ApJ, in press (astro-ph/0706.0516)\\
Smecker-Hane, T.,A., McWilliam, A., 1999, ASP Conf. Ser. 192, p.150\\
Smecker-Hane, T.,A., McWilliam, A., 2002 astro-ph/0205411\\  
Smith, E.O.; Neill, J.D.; Mighell, K.J.; Rich, R. M. 1996, AJ 111, 1596\\
%Smith, V.V. (1997) Rev. Mod. Phys. 69, 995\\            
%CC_Sept5 
Smith, V.V., Hinkle, K.H., Cunha, K., Plez, B., et al., 2002, AJ, 124, 3241 \\
%Smith, V.V., Suntzeff, N.B., Cunha, K., Gallino, R., Busso, M., Lambert, D.L., 
%Straniero, O., 2000, AJ, 119, 1239 \\
%CC_Sept5 
%Sneden, C. 1973, ApJ, 184, 839\\
%Sneden, C., Pilachowski, C., A,. \& Kraft, R., P. 2000, AJ, 120, 1014\\
Stephens, A., \& Boesgaard, A.M. 2002, AJ, 123, 1647\\
Stetson, P.B., VandenBerg, D.A., \& Bolte, M. 1996, PASP, 108, 560   \\
%Suntzeff, N.B., \& Kraft, R.P. 1996, AJ, 111, 1913\\
Tautvaisiene, G.; Wallerstein, G.; Geisler, D.; Gonzalez, G.; Charbonnel, C. 
2004, AJ, 127, 373\\
Tautvaisiene, G., Geisler, D., Wallerstein, G., Borissova, J., Bizyaev, D., 
Pagel, B.E.J., Charbonnel, C., Smith V.V. 2007, AJ, submitted \\
%Thevenin \& Idiart (1999), ApJ, 521, 753\\
Timmes, F. X.; Woosley, S. E.; Weaver, T.A. 1995, ApJS, 98, 617\\
Tinsley, B.M. 1979, ApJ, 229, 1046\\
Tolstoy, E., Irwin, M.J., Cole, A.A., Pasquini, L., Gillmozzi, R., \& Gallagher,J. S. 2001, MNRAS, 327, 918\\
Tolstoy, E., Venn, K., A., Shetrone, M., Primas, F., Hill, V., Kaufer, A., \& Szeifert, T. 2003, AJ, 125, 707 (T03)\\
Tolstoy, E.; Irwin, M. J.; Helmi, A.; Battaglia, G.; Jablonka, P.; Hill, V.; Venn, K. A.; Shetrone, M. D.; Letarte, B.; Cole, A. A.; et al. 
2004, ApJ, 617L, 119\\
Tolstoy, E. 2005, IAU Coll. 198, Near Field Cosmology with Dwarf Elliptical
Galaxies, ed. H. Jerjen \& B. Binggeli, p. 118 (T05)\\
Tsuchiya, T., Dinescu, D. I., Korchagin, V. I. 2003, ApJ 598, L29\\
Tsuchiya, T., Korchagin, V. I., Dinescu, D. I. 2004, MNRAS 350, 1141\\
Unavane, M., Wyse, R.F.G., \& Gilmore, G. 1996, MNRAS, 278, 727\\
Vanture A. D., Wallerstein, G., \& Brown, J. A. 1994 PASP 106, 835\\
Venn, K.A., Kaufer, A., McCarthy, J.M., et al. 2001, ApJ, 547, 765\\
Venn, K.A., Tolstoy, E., Kaufer, A., et al. 2003, AJ, 126, 1326\\
Venn, K.A.; Irwin, M.; Shetrone, M.D.; Tout, C.A.; Hill, 
V.; \& Tolstoy, E.  2004 AJ,   128, 1177\\
%Ventura, P., D'Antona, F., Mazzitelli, I., \& Gratton, R. 2001, ApJ, 550, 
%L65\\
Ventura, P., \& D'Antona, F. 2005, ApJ, 635, L149 \\
%Walcher, J., Fried, J.W., Burkert, A., \& Klessen, R.S. 2003, A\&A 406, 847\\  
Wallerstein, G.; Iben, I., Jr.; Parker, P.; Boesgaard, A.M.; Hale, G.M.; Champagne, A.E.; Barnes, C.A.; Kappeler, F.; Smith, V.V.; Hoffman, R.D.; et al.
1997, RvMp, 69, 995\\
%Weiss,A. \& Charbonnel,C. 2003, in IAU General Assembly, Joint Discussion 04,
%ed. F. D'Antona and R. Gratton, in press \\
White, S. D. M.; Rees, M. J. 1978, MNRAS, 183, 341\\
Willman, B. et al. 2005, ApJ 626L, 85\\
Woosley, S.E., \& Weaver, T.A. 1995, ApJS, 101, 181\\
%Yanny, B., et al., 2003, ApJ, 588, 824\\
Zinn, R. 1980, ApJS, 42, 19\\
Zinn, R. 1985, ApJ, 293, 424\\
Zinn, R. 1993, ASP Conf. Ser., 48, 38\\
Zucker, D. B.; Belokurov, V.; Evans, N. W.; Wilkinson, M. I.; Irwin, M. J.; Sivarani, T.; Hodgkin, S.; Bramich, D. M.; Irwin, J. M.; Gilmore, G.; et al.
2006a, ApJ, 643L, 103\\
Zucker, D. B.; Belokurov, V.; Evans, N. W.; Kleyna, J. T.; Irwin, M. J.; Wilkinson, M. I.; Fellhauer, M.; Bramich, D. M.; Gilmore, G.; Newberg, H. J.; 
2006b ApJ, 650L, 41\\
Zucker, D.B.; Belokurov, V.; Evans, N. W.; Gilmore, G.; Wilkinson, M. I. 2006c,
AAS, 2091, 7805\\

%\end{thebibliography}

\clearpage

\begin{figure}
\plotone{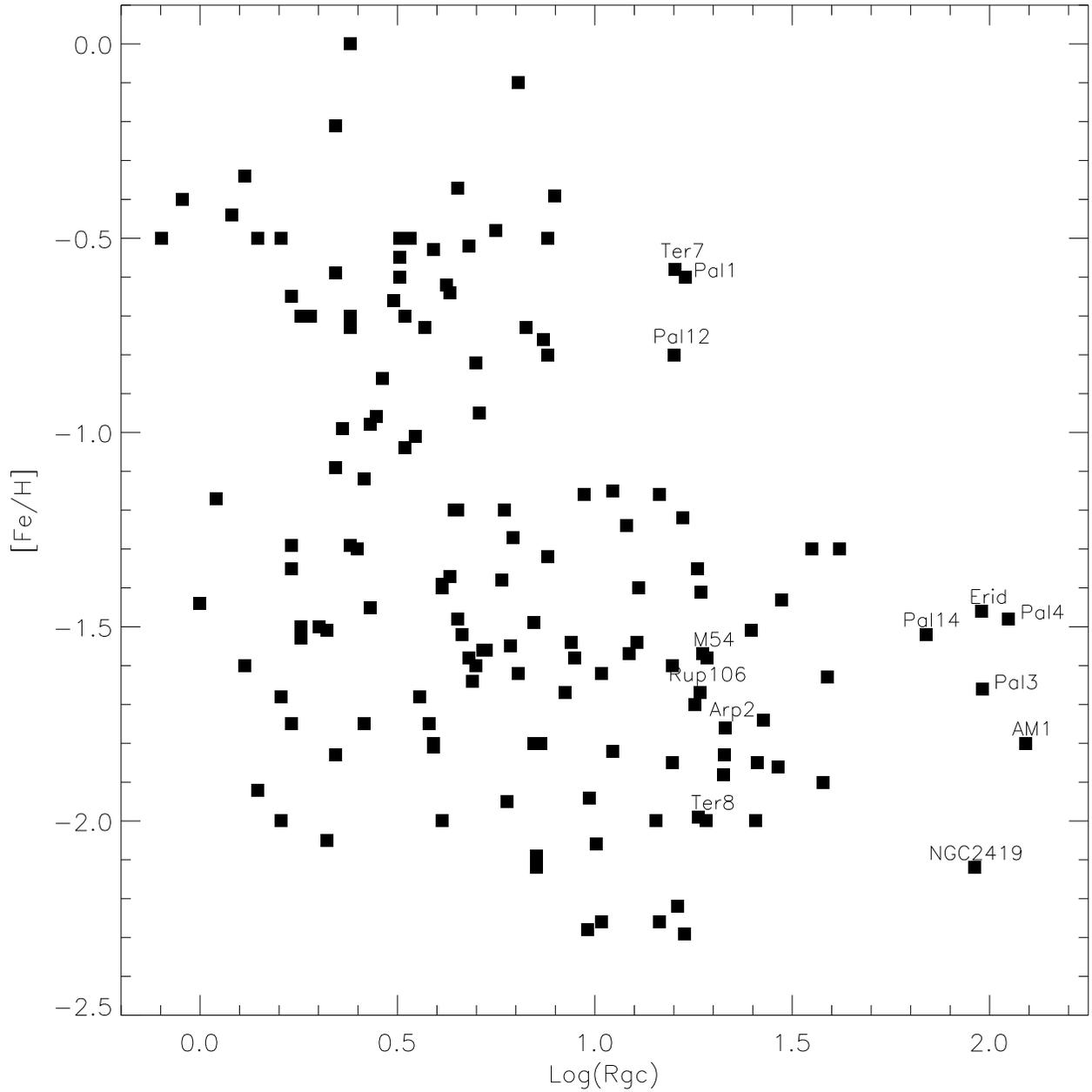}
\caption{
The metallicity of globular clusters as tabulated by Harris (2003)
plotted against the log of their distances from the Galactic Center in
kiloparecs. The break at 40 kpc is evident followed by 6 very distant
globulars located
at similar distances to the nearer dSph systems.
Various clusters are identified.
}
\end{figure}

\clearpage

\begin{figure}
\plotone{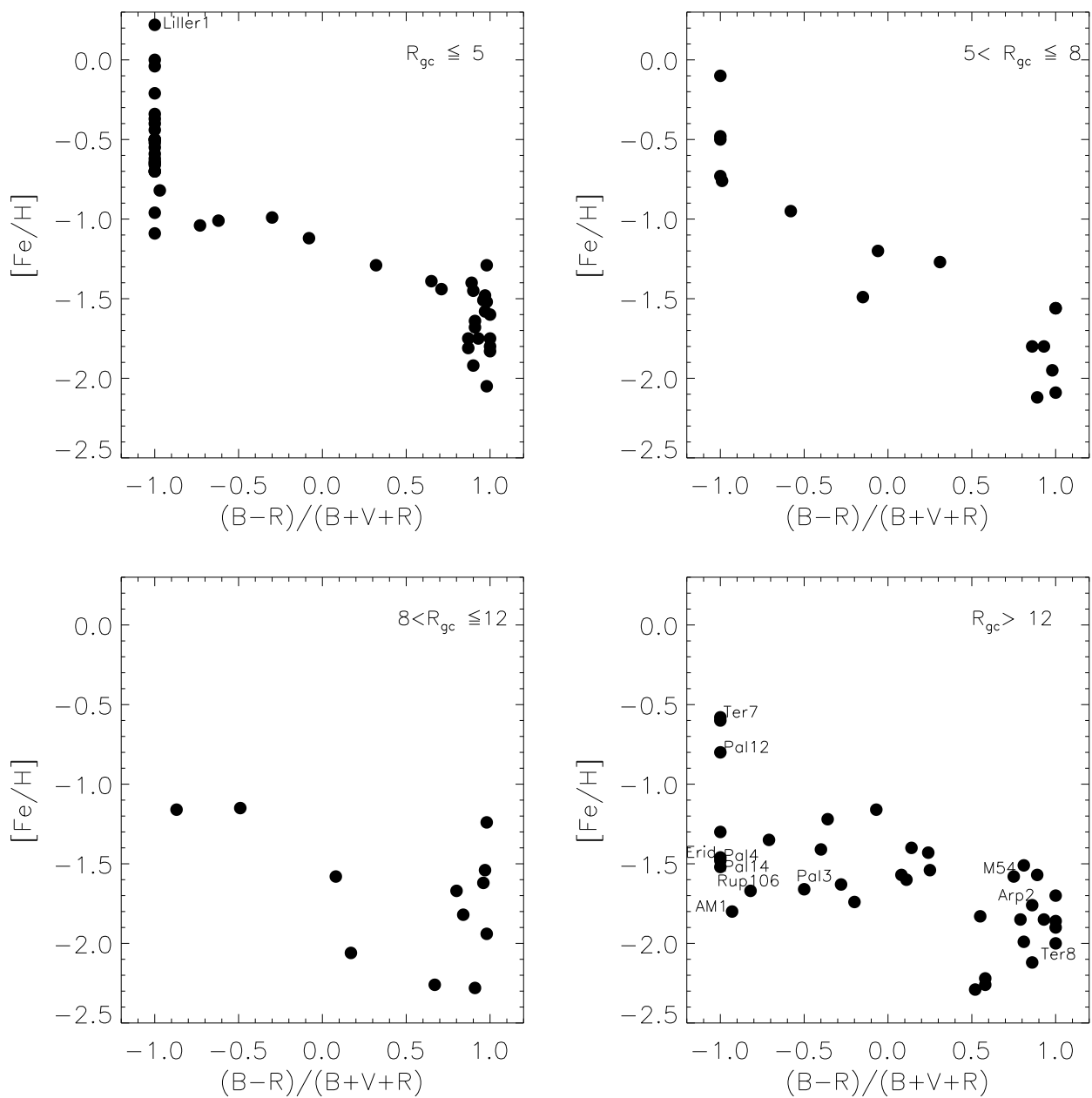}
\caption{
The dependence of metallicity on the color distribution of
horizontal branch stars for globulars divided into 4 groups according to
their distance from the Galactic Center, following SZ. Beyond 8 kpc there are no
clusters with [Fe/H] > -1.0 except for Ter 7 and Pal 12 which are
associated with the Sgr System. NGC 2419 is located in the bottom right diagram
at (0.86,-2.12).
}
\end{figure}

\clearpage

\begin{figure}
\includegraphics[scale=.80,angle=-90]{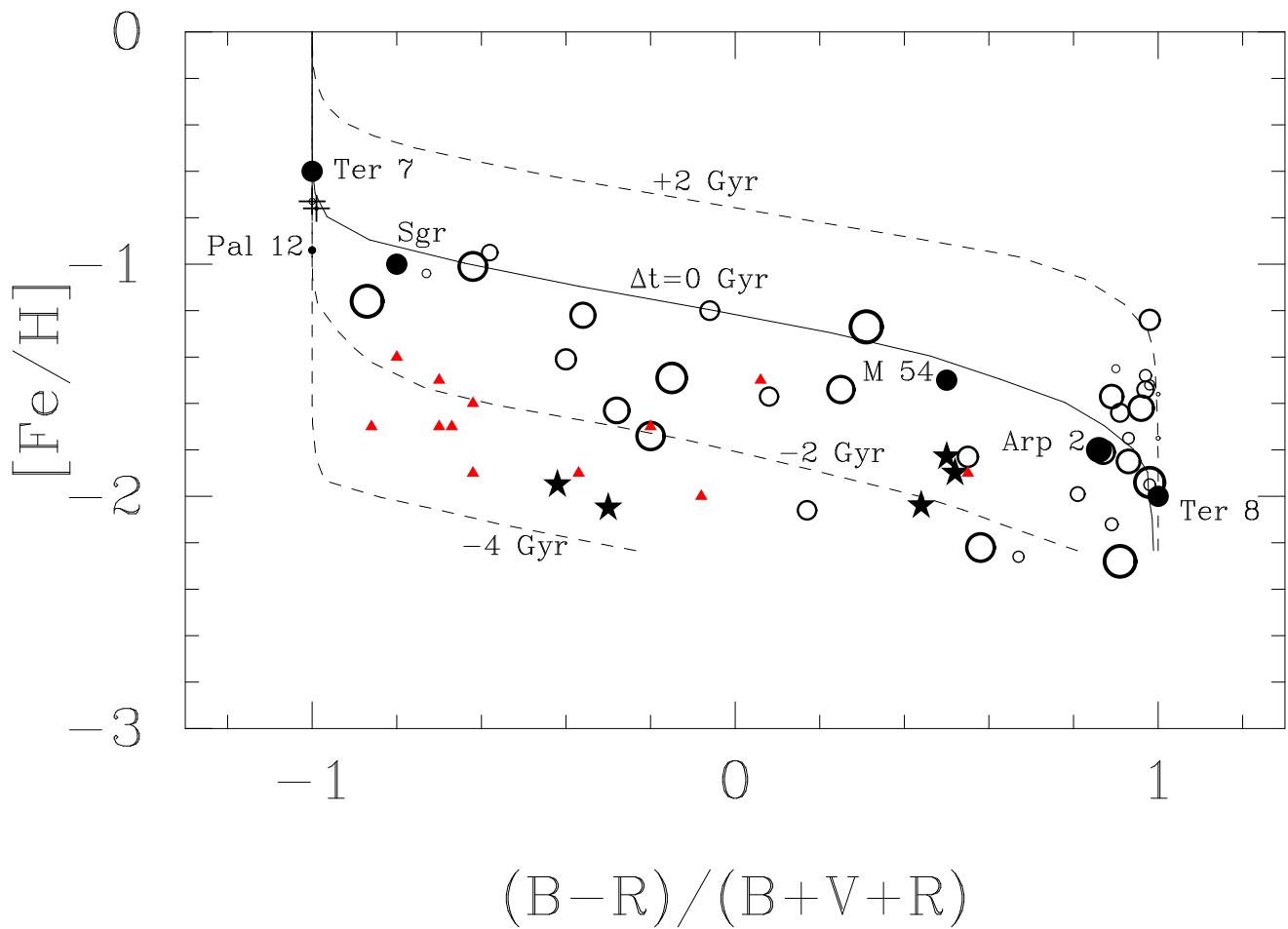}
\caption{
Metallicities as a function of HB type (Lee diagram) for
Galactic globular clusters (circles), Fornax
dwarf spheroidal clusters (star symbols) and
mean values for dwarf spheroidals (triangles). For the Galactic
globular clusters, the symbol size is proportional to the eccentricity.
Sgr's clusters are represented with filled symbols and labeled.
Equal age lines from Lee et al (2002) are also shown.
}
\end{figure}

\clearpage

\begin{figure}
\plotone{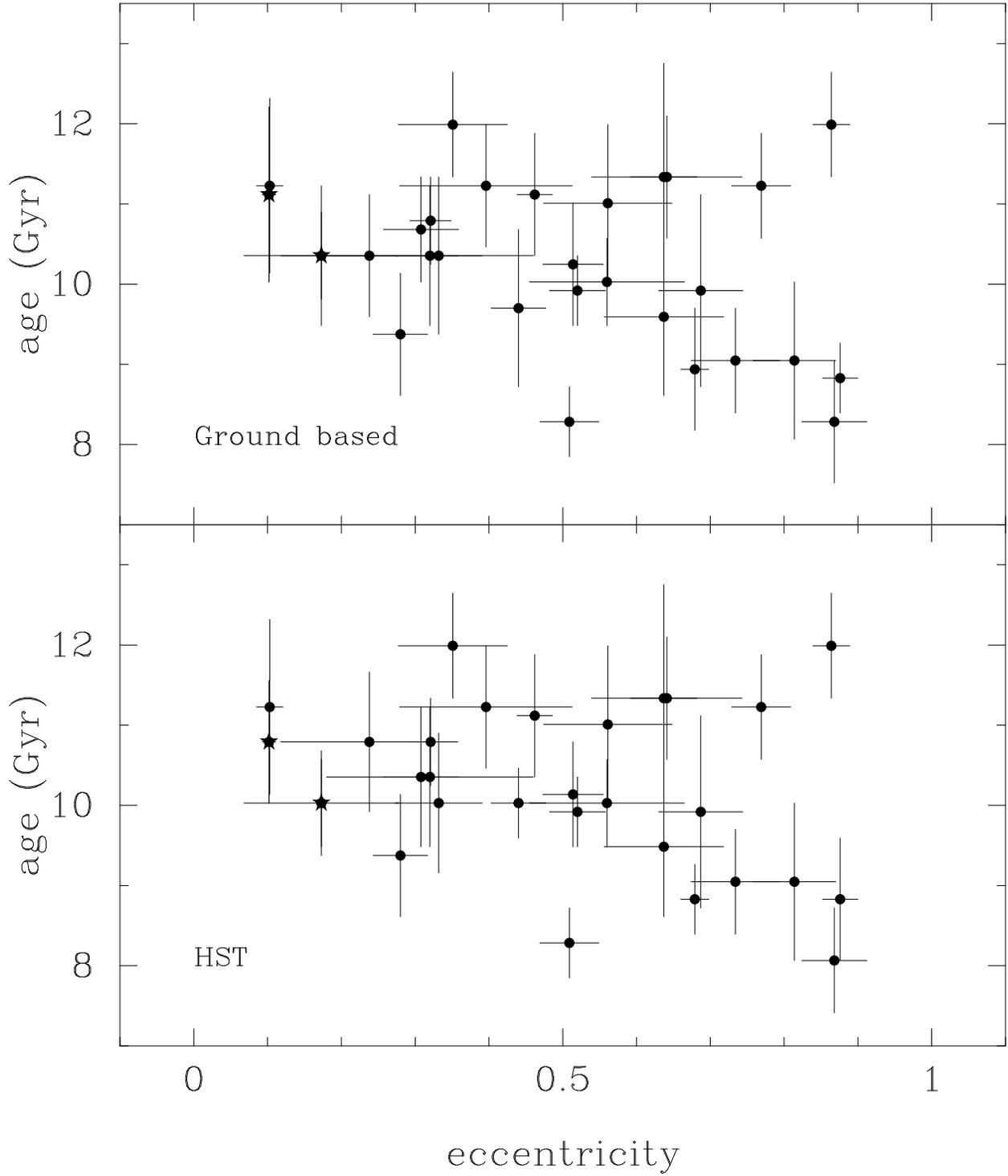}
\caption{
Ages for Galactic GCs from De Angeli et al. (2005) as a function of orbital eccentricity for two age datasets:
ground based (top) and HST (bottom). Two metal rich clusters
([Fe/H $> -0.8$) are represented with star symbols.
}
\end{figure}

\clearpage

\begin{figure}
\plotone{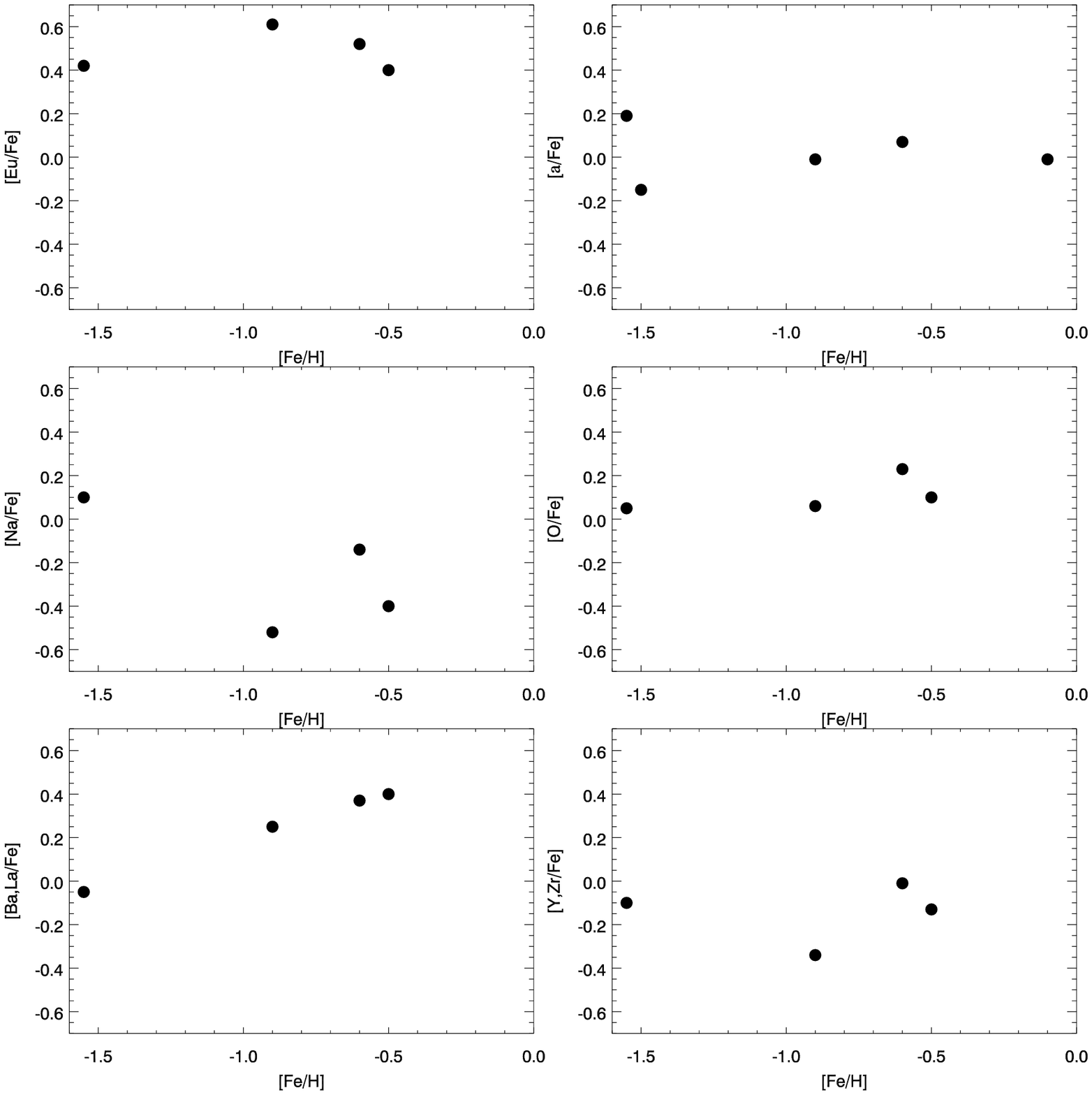}
\caption{
Trends of abundance of various elements or element
groups with \feh
in the various systems that are associated with the Sagittarius Galaxy.
}
\end{figure}

\clearpage

\begin{figure}
\plotone{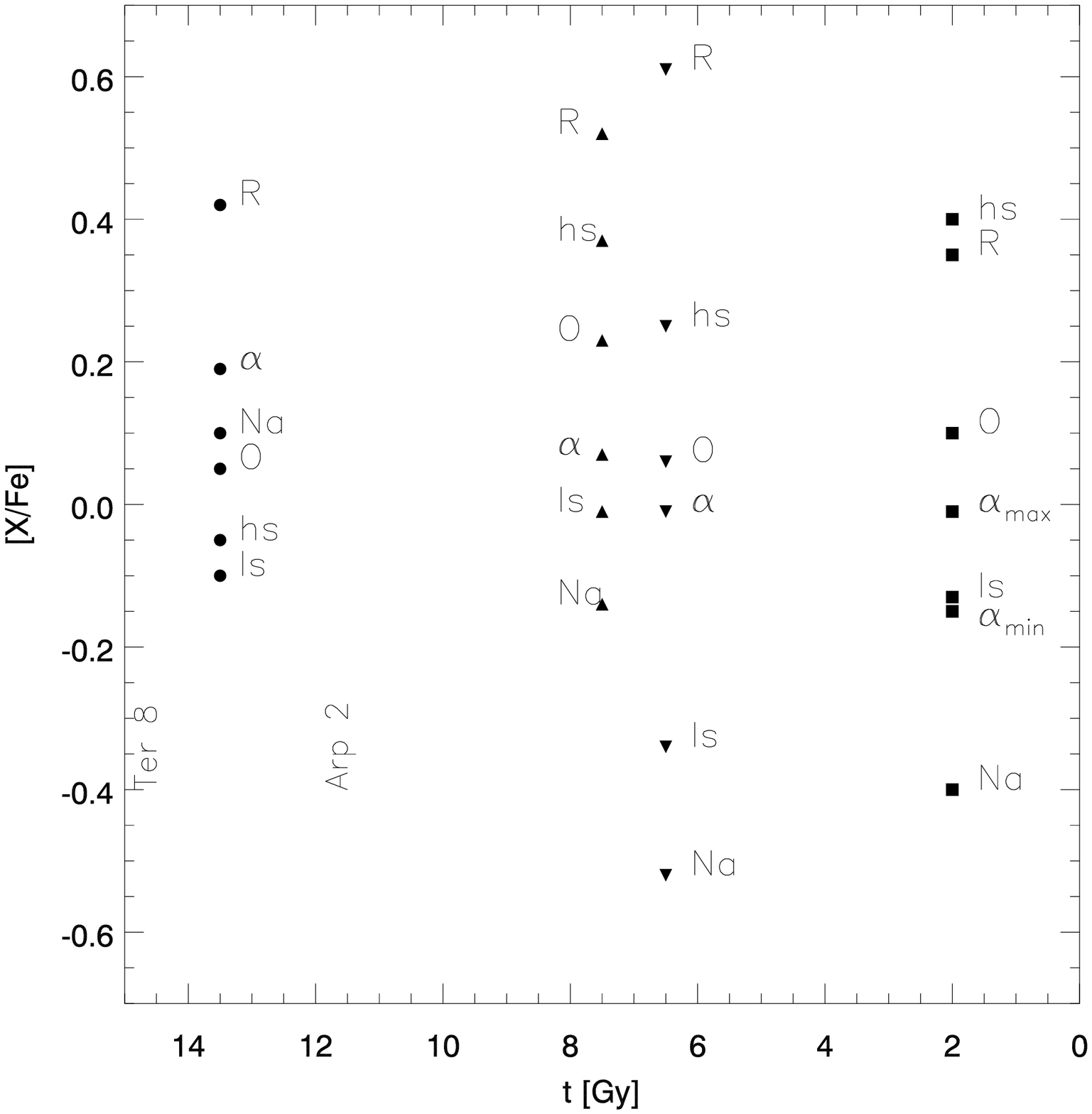}
\caption{
Trends of abundance of various elements or element
groups with age in the various systems that are associated with the Sagittarius
Galaxy.
}
\end{figure}

\clearpage

\begin{figure}
\epsscale{.80}
\plotone{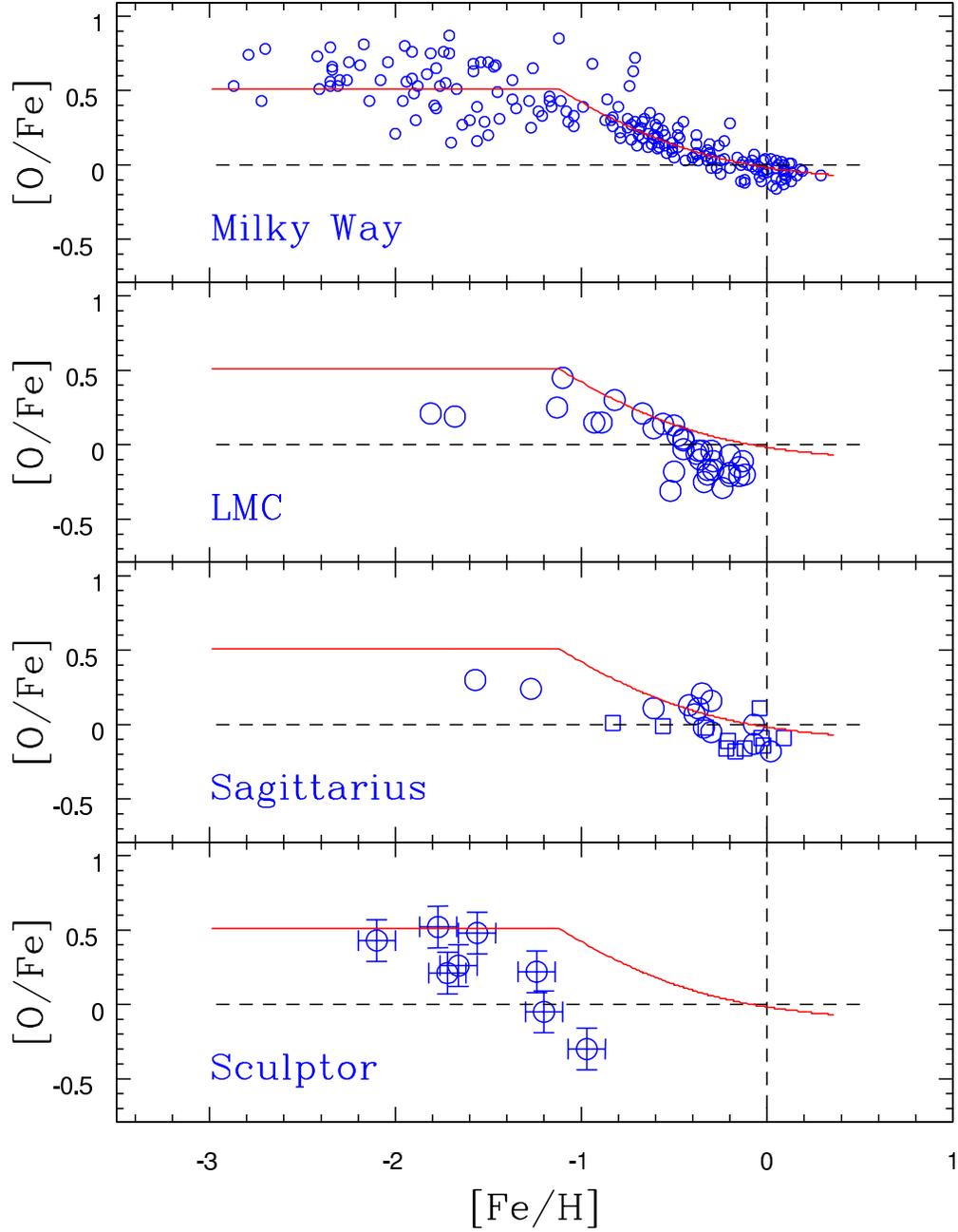}
\caption{
[O/Fe] vs. \feh for samples of stars in 4 different \gals. The solid
curve is a model fit described in the text. O \abus in the extraGalactic samples
are generally depleted with respect to their Galactic counterparts.
}
\end{figure}

\clearpage

\begin{figure}
\plotone{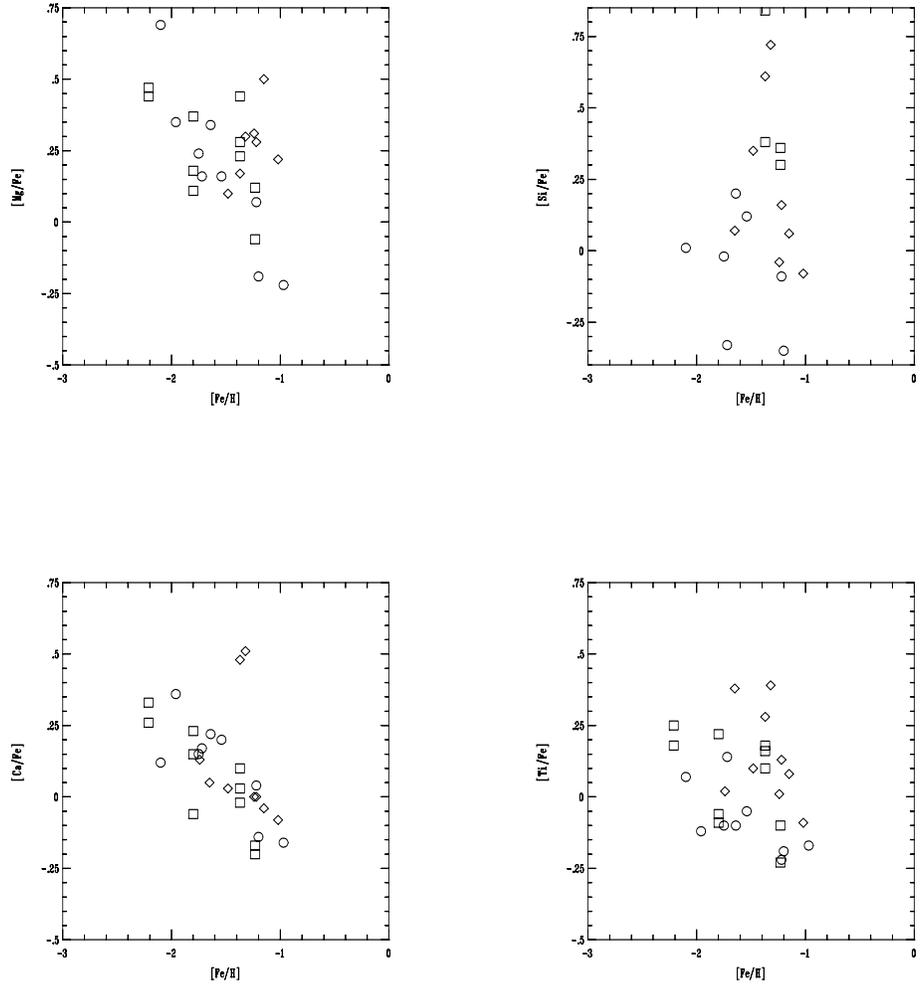}
\caption{
Mg, Si, Ca and Ti \abus vs. \feh for giants in Scl (circles) and 2
LMC samples (J05 clusters - squares, P06 field stars - diamonds).
}
\end{figure}

\clearpage

\begin{figure}
\plotone{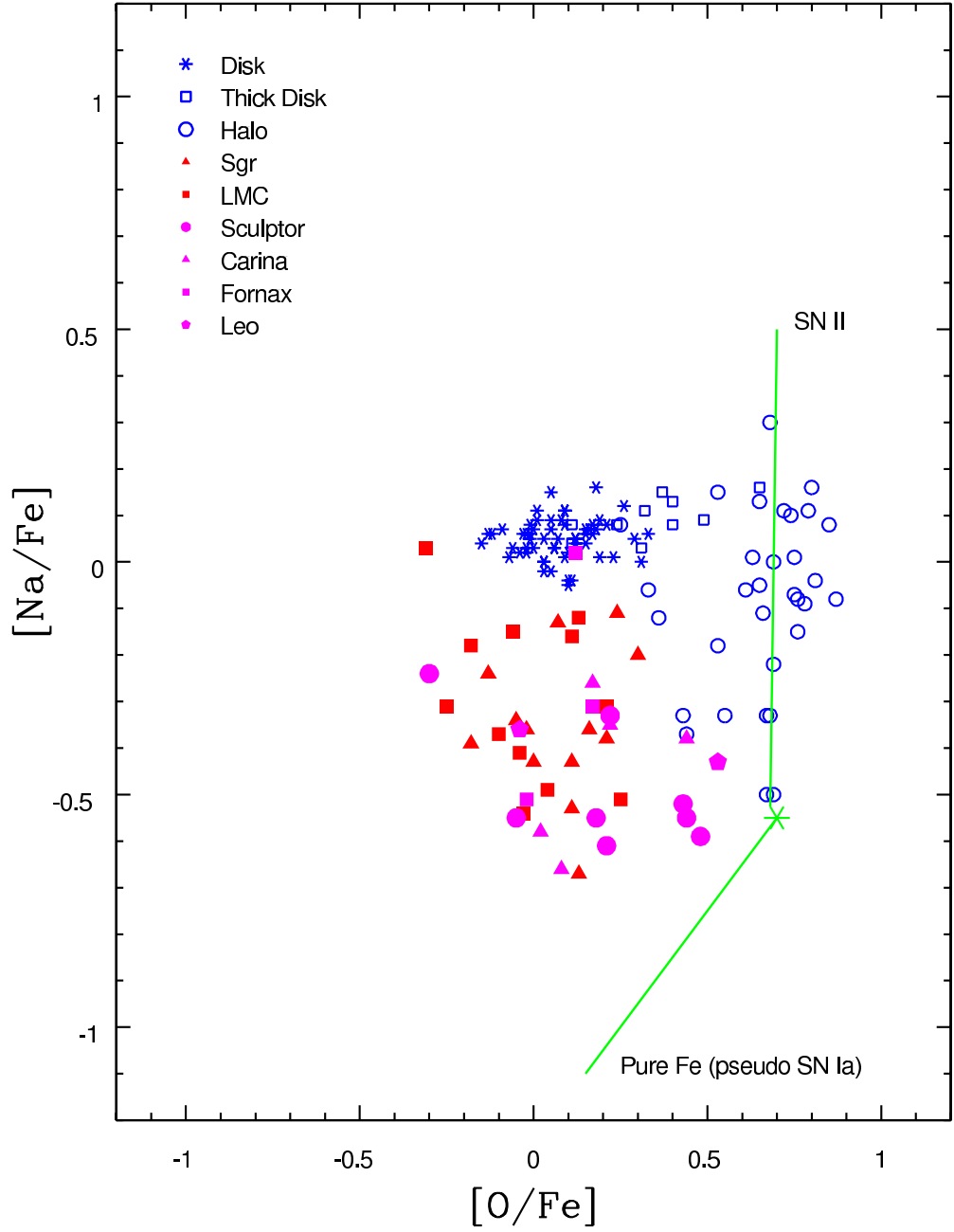}
\caption{
[Na/Fe] vs/ [O/Fe] for stars in the Galaxy (blue symbols) and
various extraGalactic samples (magenta - see key). Solid green lines represent schematic
representations of contributions expected from pure SNe II and SNe Ia.
}
\end{figure}

\clearpage

\begin{figure}
\plotone{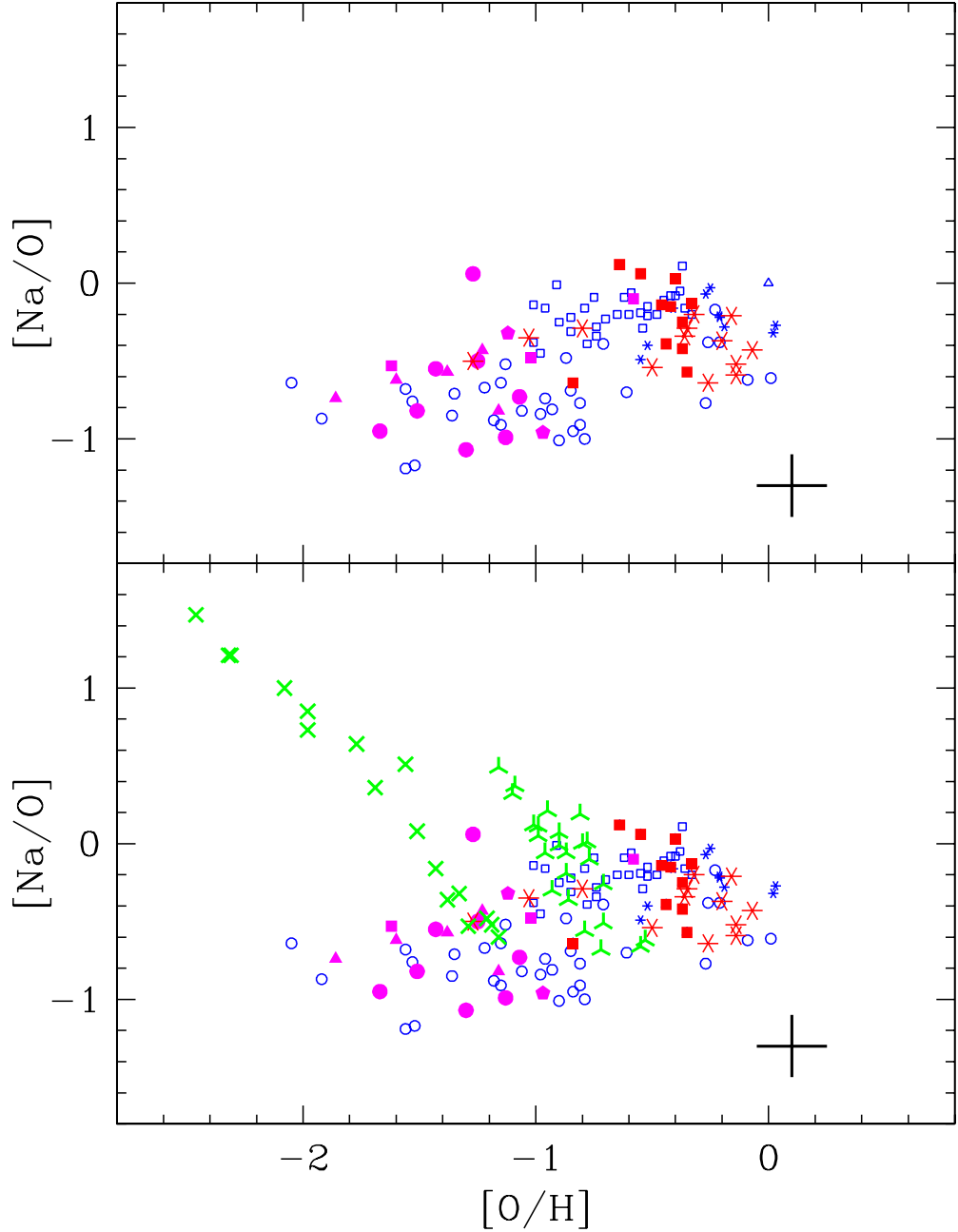}
\caption{
[Na/O] vs. [O/H] for (top) field stars in the Milky Way and dwarf
galaxies (symbols as in Figure 9) and (bottom) as in the top panel but also
including stars from two globular clusters in green: M13 (crosses) and M4
(three-pronged symbol).
}
\end{figure}

\clearpage

\begin{figure}
\plotone{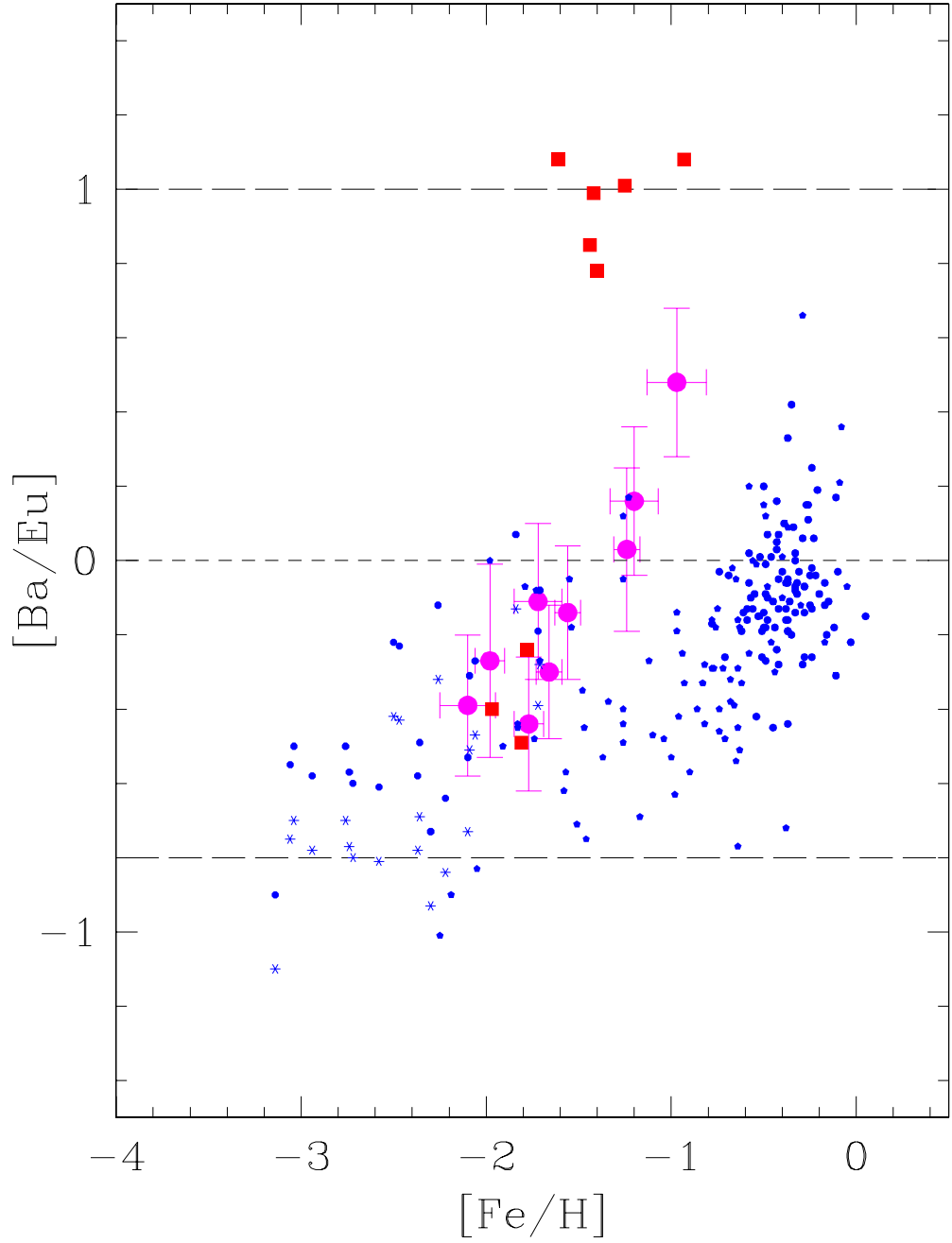}
\caption{
[Ba/Eu] vs. \feh for Galactic stars (same symbols as Figure 9),
dSph stars, Scl (large magenta circles) and $\Omega$ Cen (red squares). The
dashed lines show expected \abus for pure r-process (bottom) and pure s-process
(top).
}
\end{figure}

\clearpage

\begin{figure}
\epsscale{1.0}
\plotone{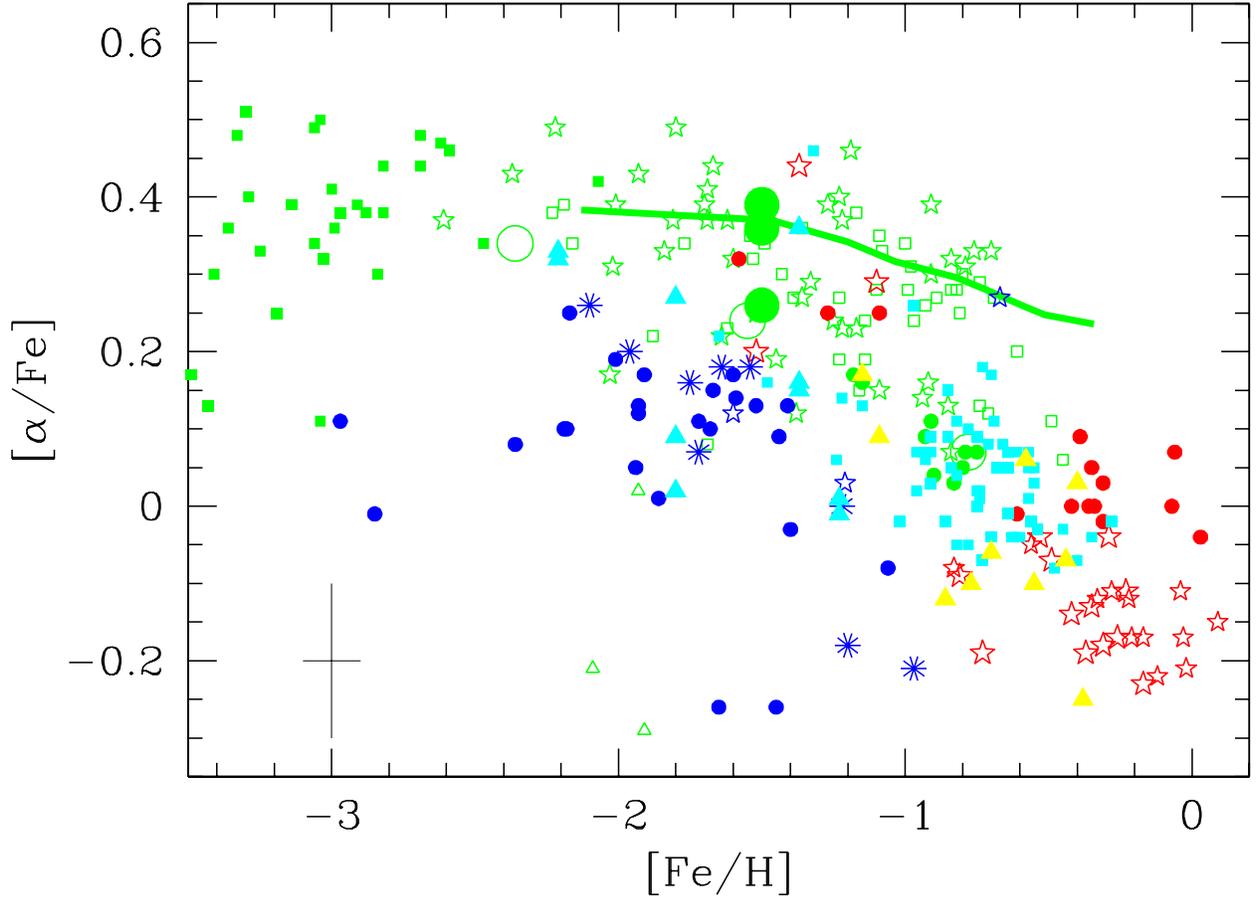}
\caption{
$[\alpha/Fe]$ vs. \feh for a variety of halo and extraGalactic samples (see text for
full explanation). Green symbols represent halo     samples, blue are for low-mass
dSphs, red for Sgr, cyan for the LMC, and yellow for dIrrs.
In detail, 
solid green curve and green stars - Gratton et al. (2003) dissipative collapse 
and accreted halo stars, respectively; small green filled circles - NS97 low \alp stars;
mean values for the three F02 components as the large green filled circles;
mean points for 3 different \met bins from Stephens and Boesgaard (2002) - 
large green open circles;
Ivans et al. (2003) - green triangles; 
Cayrel et al. (2004) - green filled squares;
Jonsell et al. (2005) - green open squares.
The blue stars are the Fornax dSph and the blue filled circles
are stars in other dSphs studied by Shetrone et al.\ (2001) and Shetrone et al.\ (2003).
Blue asterisks are from the Scl dSph studies of S03 and G05.
Red filled circles are the Sgr sample of McWilliam \& Smecker-Hane (2005) and the red stars
are the Sgr samples of Bonifacio et al.\ (2004) and Monaco et al.\ (2005).
The cyan triangles indicate the LMC clusters studied by Johnson et al.\ (2006) and the cyan
squares LMC field  stars from Pompeia et al. (2007).
Observations of stars in dwarf irregular
galaxies, including NGC 6822 (Venn et al.\ 2001), WLM (Venn et al.\ 2003) 
Sextans A (Kaufer et al.\ 2004) and IC 1613 (Tautvai\v{s}ien\.{e} et al.\ 2007),
  are indicated as the yellow filled triangles.
}
\end{figure}

\clearpage

\begin{figure}
\plotone{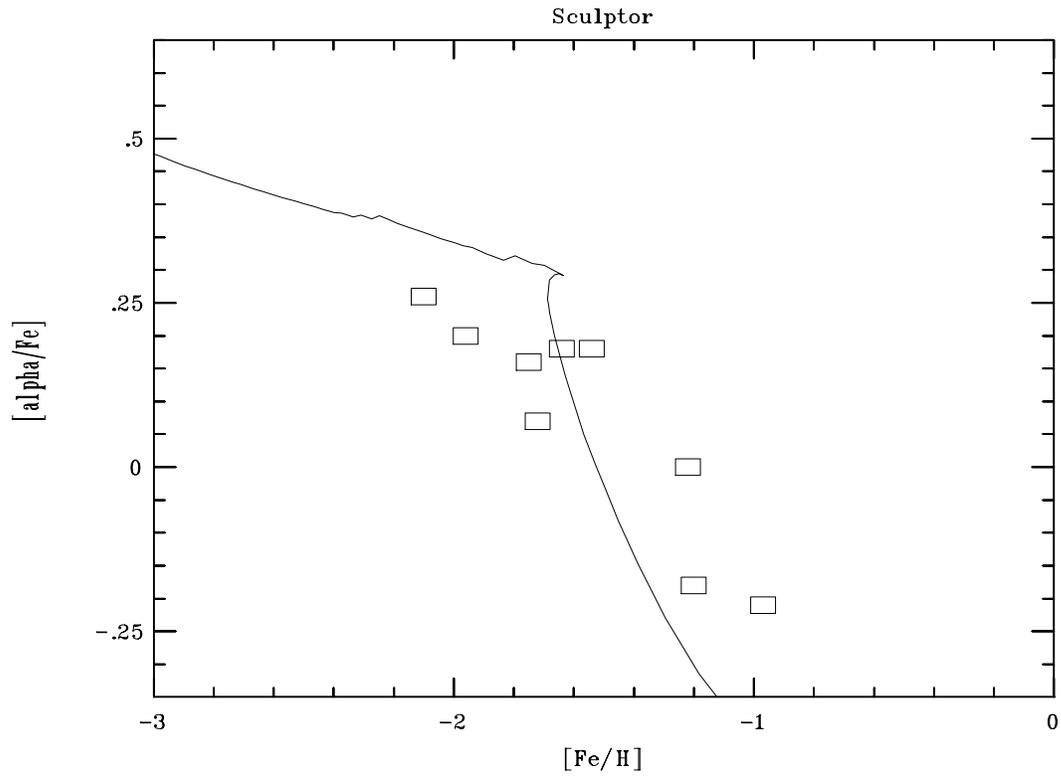}
\caption{
Figure 13: $[\alpha/Fe]$ vs. \feh for Scl dSph giants (squares) compared to
the theoretical predictions of LF04 (solid curve).
}
\end{figure}

\clearpage

\begin{figure}
\plotone{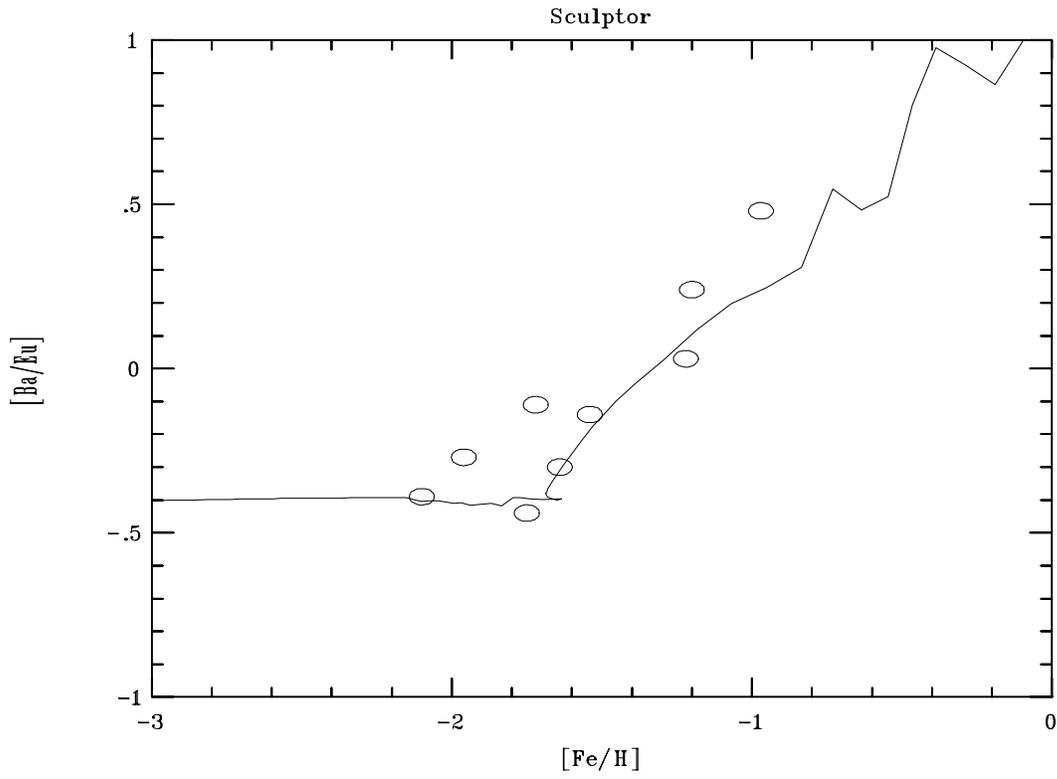}
\caption{
[Ba/Eu] vs. \feh for Scl dSph giants (squares) compared to
the theoretical predictions of Landfranchi et al.  (2006 - solid curve).
}
\end{figure}

\end{document}